\documentclass[preprint,times]{elsarticle}
\pdfoutput = 1
\journal{}
\usepackage{ecrc}
\usepackage{textcomp}
\usepackage{amsmath}
\usepackage{bm}
\usepackage{float}
\usepackage{subfigure}
\usepackage{float}

\newcommand{\nn}{{\mathbf n}}
\usepackage{xcolor}
\usepackage{soul}

\newcommand{\be}{\begin{equation}}
\newcommand{\ee}{\end{equation}}
\newcommand{\ba}{\begin{eqnarray}}
\newcommand{\ea}{\end{eqnarray}}

\newcommand{\al}{\alpha}
\newcommand{\bt}{\beta}
\newcommand{\gm}{\gamma}

\newcommand{\kp}{\kappa}

\volume{00}

\jid{}
\firstpage{1}
\journalname{}
\runauth{}

\begin{document}
\begin{frontmatter}

\title{A micromechanical analysis of intergranular stress corrosion cracking of an irradiated austenitic stainless steel}

\author[a]{D. Liang}
\author[a,cor1]{J. Hure}
\author[a]{A. Courcelle}
\author[b]{S. El Shawish}
\author[a]{B. Tanguy}
\cortext[cor1]{Corresponding author: jeremy.hure@cea.fr}
\address[a]{Universit\'e Paris-Saclay, CEA, Service d'\'Etude des Mat\'eriaux Irradi\'es, 91191, Gif-sur-Yvette, France}
\address[b]{Jo\v zef Stefan Institute, SI-1000, Ljubljana, Slovenia}

\begin{abstract}
  Irradiation Assisted Stress Corrosion Cracking (IASCC) is a material degradation phenomenon affecting austenitic stainless steels used in nuclear Pressurized Water Reactors (PWR), leading to the initiation and propagation of intergranular cracks. Such phenomenon belongs to the broader class of InterGranular Stress Corrosion Cracking (IGSCC). A micromechanical analysis of IGSCC of an irradiated austenitic stainless steel is performed in this study to assess local cracking conditions. A 304L proton irradiated sample tested in PWR environment and showing intergranular cracking is investigated. Serial sectioning, Electron BackScatter Diffraction (EBSD) and a two-step misalignment procedure are performed to reconstruct the 3D microstructure over an extended volume, to assess statistically cracking criteria. A methodology is also developed to compute Grain Boundary (GB) normal orientations based on the EBSD measurements. The statistical analysis shows that cracking occurs preferentially for GB normals aligned with the mechanical loading axis, but also for low values of the Luster-Morris slip transmission parameter. Micromechanical simulations based on the reconstructed 3D microstructure, FFT-based solver and crystal plasticity constitutive equations modified to account for slip transmission at grain boundaries are finally performed. These simulations rationalize the correlation obtained experimentally into a single stress-based criterion. The actual strengths and weaknesses of such micromechanical approach are discussed.
\end{abstract}


\end{frontmatter}

\section{Introduction}
Failures of austenitic stainless steels Baffle-to-Former Bolts (BFB) in the core of nuclear Pressurized Water Reactors (PWR) have been reported since the 80's as a result of an InterGranular Stress Corrosion Cracking (IGSCC) degradation phenomenon \cite{ANDRESEN2019380}. As these materials are mainly immune to SCC in aqueous environment, the phenomenon is referred to as Irradiation Assisted Stress Corrosion Cracking (IASCC) as core materials in PWR are heavily irradiated with neutrons \cite{CHOPRA2011235}. \textcolor{black}{Neutron irradiation leads to the formation of irradiation defects as a result of the recombination of point defects generated by ballistic interactions between neutrons and the atoms of the material \cite{NORDLUND2018450}. The main irradiation defect observed in austenitic steels in PWR conditions ($\sim 300^{\circ}$C) is Frank dislocation loop, but other irradiation defects are also observed such as Helium bubbles and precipitates \cite{tan2016}. In addition, irradiation induces depletion of Cr and enrichment of Ni at grain boundaries.} Understanding and predicting IGSCC of irradiated austenitic stainless steels, but also of unirradiated alloys such as Nickel-based materials, still remains a challenge \cite{2942,LIU20181}. The use of ion-irradiated materials (\textit{e.g.}, protons \cite{WAS2002198}) in the last decades, relieving the burden of dealing with neutron irradiated radioactive samples, allows a deeper understanding of the physical mechanisms involved in IGSCC of irradiated austenitic stainless steels \cite{JIAO2008203,GUPTA201682}. Both radiation-induced hardening and radiation-induced segregation at Grain Boundaries (GB) are observed and act synergistically to favor IGSCC \cite{BUSBY200220,doi:10.1080/18811248.2004.9726351}. Moreover, irradiation induces changes in intragranular deformation mechanisms and dislocation channelling is often reported for irradiated materials. For low applied strain, which is relevant for IGSCC cracking initiation, deformation occurs only in narrow bands of about tens of nanometers width as a result of the presence of irradiation defects \cite{LEE20013269,LEE20013277}. The impingement of these dislocation channels on GB has been proposed to lead to high local stresses responsible for GB cracking \cite{MCMURTREY2015305}. More precisely, \textcolor{black}{cracking has been shown to correlate with the discontinuity of traces of dislocation channels across GB on specimen surfaces (see, \textit{e.g.}, \cite{STEPHENSON2016214})}. The presence of high local stresses at the impingement of an interrupted dislocation channel on a GB has been experimentally demonstrated through residual stress measurements \cite{JOHNSON201687}, and in more details though the use of Molecular Dynamics (MD) \cite{MCMURTREY2015305,JOHNSON201687} or Finite Element (FE) simulations \cite{sauzay}. Experiments using heavy-ion irradiated samples, leading to less localization than proton irradiation, have however questioned this mechanism as cracking initiation is also observed \cite{GUPTA201845}. In addition, GB well oriented with respect to the mechanical loading direction have been found to be more prone to cracking \cite{west}, emphasizing the role of intergranular normal stresses. \textcolor{black}{Similar results have been obtained on irradiated austenitic steels in Super Critical Water (400$^{\circ}$C) \cite{WEST2011142} as well as a correlation between cracking and high Schmid factor mismatches. These observations have been rationalized through a Schmid-Modified Grain Boundary Stress (SMGBS) model.} \textcolor{black}{The experimental determinations of local cracking conditions have been mostly made using 2D measurements (except for example in \cite{JOHNSON2019166}) such as traces of GB and dislocation channels on specimen surfaces. However 3D information about the microstructure are required to accurately compute GB normals, SMGBS model or slip transmission criteria.} \textcolor{black}{This may be done for selected cracks through ionic milling, as performed in \cite{STRATULAT2014428} to assess SMGBS model on the IGSCC of a thermally sensitized austenitic steel.} Experimental techniques are available to obtain \textcolor{black}{full }3D microstructures, either destructive such as serial sectioning coupled with 2D Electron BackScatter Diffraction (EBSD) \cite{LIU2018290} or non destructive such as Diffraction Contrast Tomography (DCT) \cite{King382}. Both techniques require dedicated procedures to avoid reconstruction artefacts, but provide valuable informations regarding grain shapes / crack patterns. Up to now, to the authors' knowledge, no such analysis has been performed in the context of IGSCC of irradiated austenitic stainless steels to assess in more details local cracking conditions.

Theoretical and numerical modelling of IGSCC have been proposed in the literature. Using synthetic or realistic polycrystalline aggregates along with crystal plasticity constitutive equations, distributions of intergranular stresses have been determined for unirradiated and irradiated austenitic stainless steels \cite{GONZALEZ201449,HURE2016231}. This approach requires estimations of GB strength to predict IGSCC which can now be assessed through the development of microtestings \cite{STRATULAT20169}. However, experimental data remains scarce, especially for irradiated materials. From a numerical perspective, cohesive zones modelling can be used to predict both initiation and propagation of intergranular cracks \cite{MUSIENKO20093840,SIMONOVSKI2015139}. The results of these simulations are heavily dependent on the accuracy of the crystal plasticity constitutive equations, which is still an ongoing work for irradiated stainless steels. In particular, reproduction of dislocation channelling phenomenon is still a challenge within the crystal plasticity framework, although significant advances have been made recently regarding modelling of strain localization at the intragranular scale \cite{SCHERER2019103768,MARANO2019262}. \textcolor{black}{All the aforementioned numerical studies account only for the anisotropic mechanical behavior of grains through the use of phenomenological crystal plasticity constitutive equations \cite{ROTERS20101152}. Intergranular stresses arise mainly as a result of deformation mismatches, as no additional constraint is imposed at GB. As an example, plastic slips can take arbitrary values close to a GB, which is not in agreement with experimental observations such as GB pile-up. A model has been proposed recently \cite{HAOUALA2020102600} to handle that issue by combining standard crystal plasticity equations with slip transmission criteria. This model is simple to implement, and has been shown to be able to reproduce phenomenon associated with dislocation pile-up at GB such as Hall-Petch effect.} Despite the current limitations of polycrystalline aggregates simulations to predict IGSCC, such simulations are nevertheless required to identify gaps where modelling efforts should be put. To the authors' knowledge, no crystal plasticity simulations have been reported on realistic microstructure for the IGSCC of irradiated austenitic steels with direct comparisons to cracking initiation experimental data.

Based on this literature review, the objectives of this study are twofold. The first one is to assess local cracking conditions based on a 3D microstructure obtained on an irradiated austenitic stainless steel tested in PWR environment and exhibiting intergranular cracks. The second objective is to perform crystal plasticity simulations based on the 3D microstructure to assess the strengths and weaknesses of such micromechanical approach, as well as directions for future researches. The paper is organized as follows: in Section~\ref{sec2}, the experimental characterization is detailed, including the description of the material, the reconstruction of the 3D microstructure, as well as the analysis of local cracking conditions. Numerical simulations are detailed in Section~\ref{sec3}. The experimental and numerical results are finally discussed in Section~\ref{sec4}.

\section{Experimental characterization} 
\label{sec2}
\subsection{Material}

The material is a Solution Annealed (1050$^{\circ}$C / 30min followed by water quench) 304L austenitic stainless steel (18.75\% Cr, 8.55\% Ni, 0.02\% Mo, 0.45\% Si, 1.65\% Mn, 0.012\% C, wt), with a mean grain size of $27\mu$m. \textcolor{black}{Solution annealing allows to bring back (intergranular) carbides into solid solution - hence reducing the susceptibility to IGSCC -  and is the heat treatment performed on the materials used for PWRs baffles and formers.} Flat tensile specimens (2mm x 2mm x 18mm gauge length) have been sampled using electrical discharge machining, and mirror polished on one side. The last polishing step is a vibratory polishing with colloidal silica solution ($0.05\mu$m, pH=7) to remove any surface hardening from previous polishing steps.

\begin{figure}[H]
\centering
\subfigure[]{\includegraphics[height = 4.6cm]{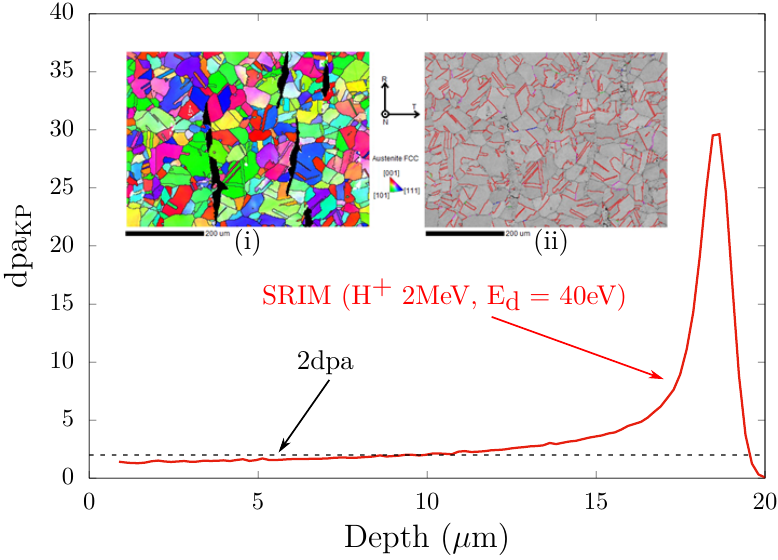}}
\hspace{1cm}
\subfigure[]{\includegraphics[height = 4.4cm]{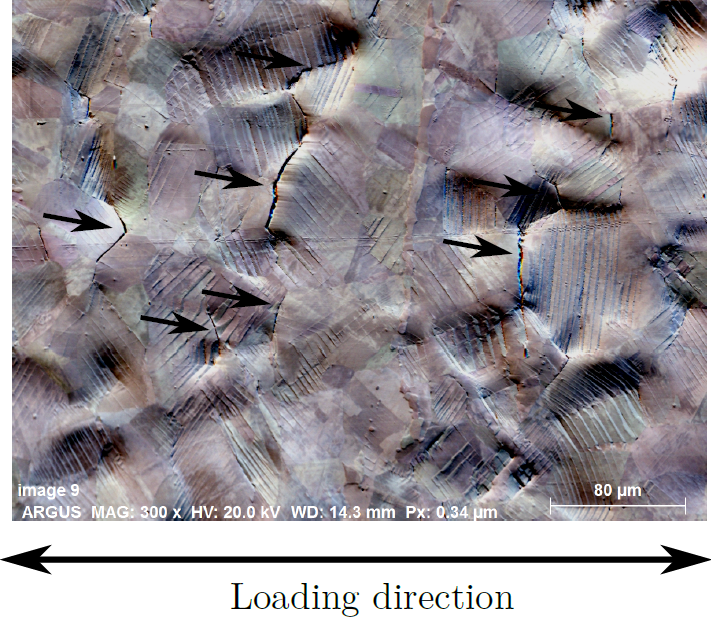}}
\caption{(a) Dose profile along the thickness of the samples computed using SRIM-2013 software (2MeV $\mathrm{H^+}$, $4.75\ 10^{19} \mathrm{H^+.cm^{-2}}$, displacement energy of $E_d=40\mathrm{eV}$, Kinchin-Pease approximation \cite{STOLLER201375}). Inset: EBSD maps showing (i) the FCC crystallographic orientations and (ii) the presence of ferrite (in black), as well as the $\Sigma_3$ GB. (b) Forward Scatter Detector (FSD) SEM image of the sample surface after the \textcolor{black}{SSRT} showing intergranular cracking afer 4\% plastic strain in PWR environment}
\label{fig1}
\end{figure}

The samples have been irradiated with 2MeV protons at the Michigan Ion Beam Laboratory (MIBL) at a temperature of 350$\pm10^{\circ}$C. The mean flux was $1.39\ 10^{14} \mathrm{H^+.cm^{-2}.s^{-1}}$, and the irradiation time $95$h. The irradiation dose profile - quantified with displacements per atoms (dpa) \cite{STOLLER201375} to compare with irradiations with different particles - is shown in Fig.~\ref{fig1}a: only the first 20$\mu$m of the samples are irradiated, with an approximately constant irradiation dose of about 2dpa over the first 10$\mu$m. A tensile sample was subjected to a Slow Strain Rate Test (SSRT) at a strain rate of $5.10^{-8}\mathrm{s^{-1}}$ up to 4\% plastic strain in PWR environment (340$^{\circ}$C, deoxygenated water with 1000ppm B, 2ppm Li, 25-35cc $\mathrm{H_2}$). After the \textcolor{black}{SSRT}, Scanning Electron Microscope (SEM) observations have confirmed the presence of intergranular cracking, as shown on Fig.~\ref{fig1}b. Details about the material, the irradiation conditions and the SCC test can be found in \cite{GUPTA201845}. This tensile sample was used to assess local conditions for intergranular cracking of irradiated austenitic steels, which first requires the reconstruction of the 3D microstructure.

\subsection{Microstructure}

The typical surface cracks density of the sample has been determined in \cite{GUPTA201845} to be about $300\mathrm{mm^{-2}}$. In order to get statistically relevant results for the local cracking conditions, the 3D microstructure has to be reconstructed on an area containing few tens of cracks at least, \textit{e.g.}, a typical surface of about 1mm$^2$. Different techniques have been proposed in the literature for this purpose. Serial-sectioning FIB tomography is rather limited to small volumes - typically $(10 \mu\mathrm{m})^3$ - due to milling rate limitations \cite{LOZANOPEREZ201278}. DCT has the advantages to be non destructive and \textcolor{black}{to allow large scan volumes}, but the detection of intergranular cracks might be difficult \cite{King382}. Therefore, the technique used in the following combined serial-sectioning polishing and 2D EBSD measurement, as for example done in \cite{LIU2018290}. \textcolor{black}{Note that an alternative technique has been proposed recently where serial-sectioning polishing is replaced by Broad-Ion-Beam (BIB) milling \cite{WINIARSKI201752}, that also allows to assess large volumes.}

\subsubsection{3D EBSD}

\begin{figure}[H]
\centering
\subfigure[]{\includegraphics[height = 4.6cm]{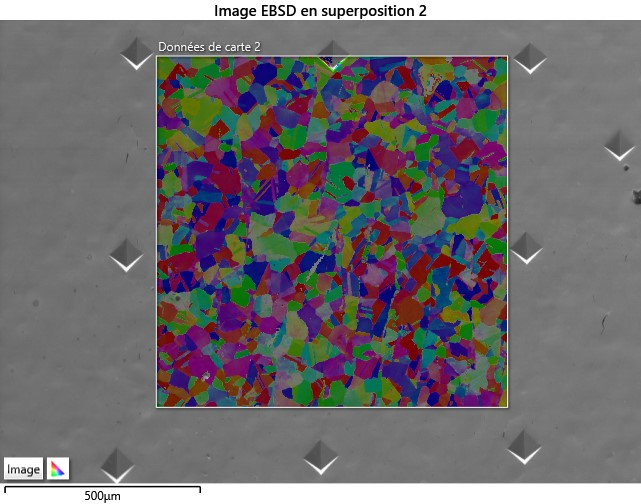}}
\hspace{1cm}
\subfigure[]{\includegraphics[height = 4.4cm]{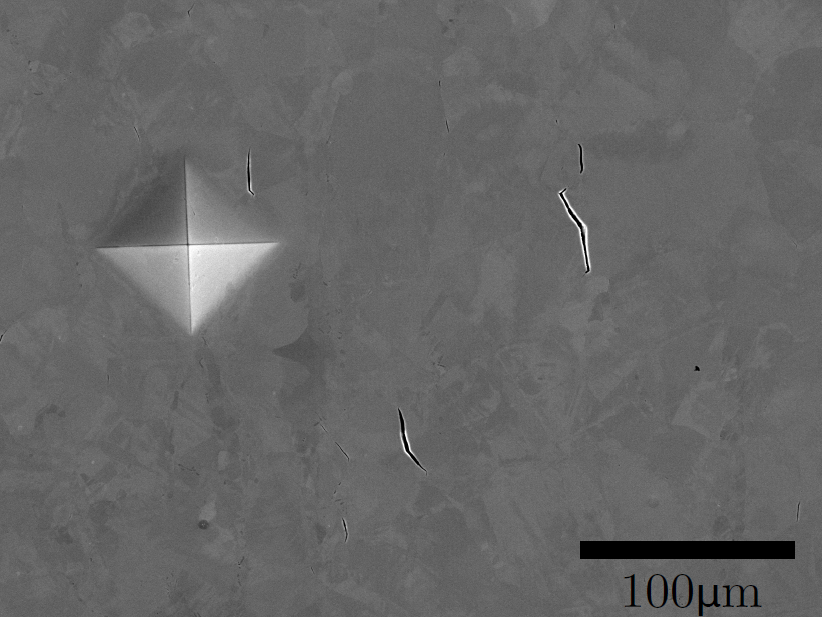}}
\caption{(a) SEM observation (Secondary Electron (SE) mode, 20kV, 70$^{\circ}$ tilt) of the Vickers indents at the surface of the sample and location of the EBSD map (b) SEM observation of intergranular cracks (SE mode, 20kV, 0$^{\circ}$ tilt)}
\label{fig2}
\end{figure}

A portion of the sample has been mounted into a conductive resin. Vickers indents identify an area of 1mm x 1mm, as shown Fig.~\ref{fig2}a. These indents are also used to measure the thickness removal after each polishing step and help to align the SEM observations, as detailed below.

The experimental methodology is as follows. Polishing is performed on an automatic polishing machine using colloidal silica solution ($0.03\mu$m) to remove about $1\mu$m after each step. The sample is then cleaned up to remove the polishing solution, and the sizes of all indents are measured to estimate the thickness removed locally\footnote{As the depth of the initial Vickers indents is smaller than the total thickness removed, new indents are regularly performed at the same locations than the previous ones.}. The use of an automatic polishing machine on a sample placed into a mounting resin allowed to keep the parallelism of the consecutive surfaces, with a typical angle lower than 0.1$^{\circ}$ \textcolor{black}{(Appendix~A, Fig.~\ref{figAAA0}a)}. SEM observations of the area identified by the Vickers indents are then performed under Secondary Electron (SE) mode to locate intergranular cracks: 20 low magnification images are taken, as shown on Fig.~\ref{fig2}b, to ensure a sufficient resolution for crack detection, and image stitching is finally used to obtain a high-resolution image over the complete surface. EBSD maps have been performed on a JEOL IT300 SEM with a tungsten filament equipped with an OXFORD EBSD detector. All EBSD maps are acquired under the same conditions at 20kV with high current. The sample is tilted at 70$^{\circ}$ with respect to the beam axis. Diffraction patterns are indexed with the AZTEC software, leading to the crystallographic phases (FCC or BCC) and orientations (through the three Euler angles $\phi_1, \Phi, \phi_2$) using a measurement step of $2\mu$m. These steps are finally repeated down to a thickness of $20\mu$m, which corresponds to the typical grain size of the material and to the thickness of the irradiated layer. The surface preparation leads to rather good indexation ratio, still the software can not find the crystallographic phases and orientations for about 2\% of the measurement points. Moreover, as shown on Fig.~\ref{fig1}a, ferrite is present in the material (about 3\% of the EBSD measurements correspond to Body Centered Cubic (BCC) phase). As numerical simulations presented in Section~\ref{sec3} and based on the experimental results require only FCC phase with full crystallographic orientations fields, the software \texttt{MTEX} \cite{bachmann2010} is used to remove the BCC phase and fill the unindexed points with FCC crystallographic orientations of the neighbouring points. \textcolor{black}{These corrections are not expected to influence the results presented hereafter as the amount of ferrite is low ($\sim 3\%$). Moreover, no crack was observed at BCC-FCC interfaces, consistently with observations on IGSCC of unirradiated austenitic stainless steels \cite{couvant2005}.}

\begin{figure}[H]
\centering
\includegraphics[height = 5.5cm]{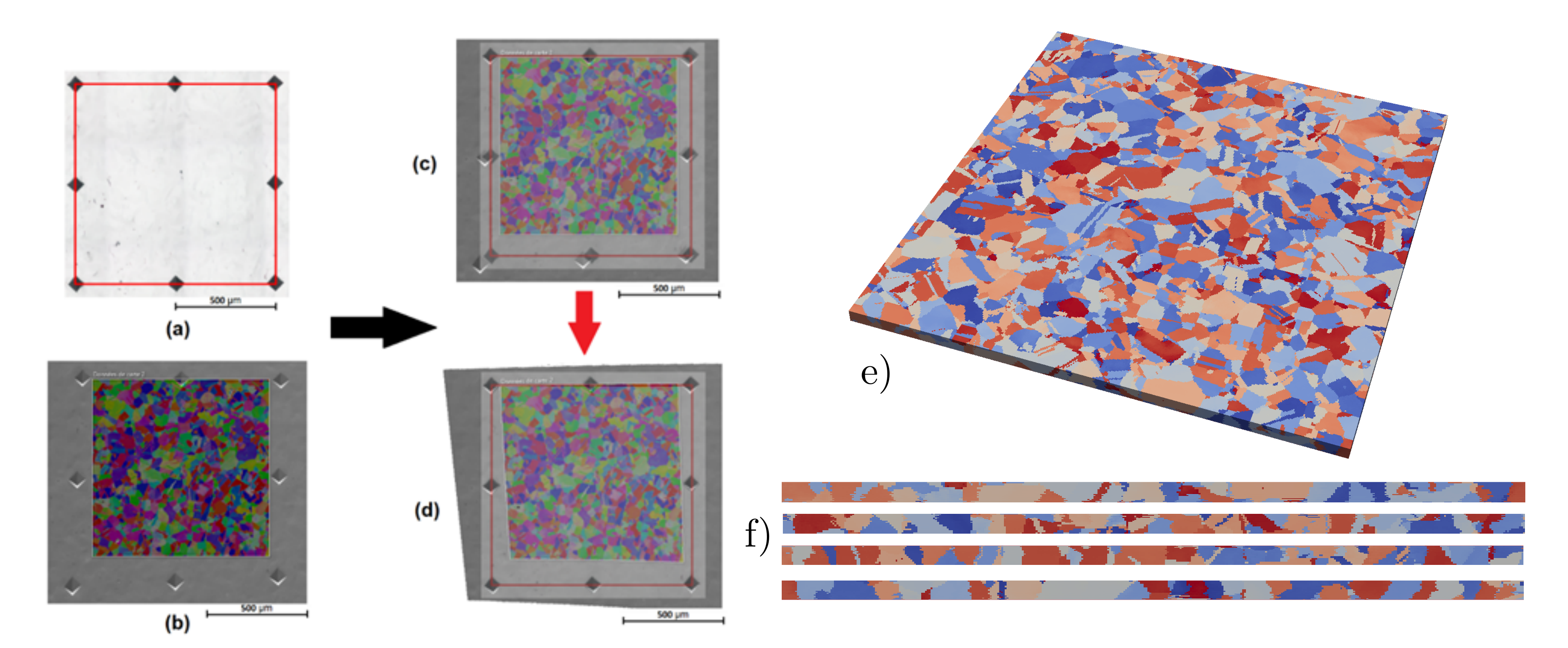}
\caption{(a) Identification of the area with Vickers indents (b) EBSD map (c,d) Correction of geometrical artefacts due to the SEM configuration used for EBSD measurements (e) 3D reconstructed microstructure: each color corresponds to a crystallographic orientation (f) Cross-sections of the 3D microstructure (see Appendix~A for cross-sections without corrections of the geometrical distortions)}
\label{fig3}
\end{figure}

\textcolor{black}{As detailed in previous studies (see \cite{PIRGAZI2019223} and references therein)}, the main challenge of 3D EBSD through serial sectioning lies in the reconstruction procedure. First, each 2D EBSD map is potentially affected by at least two effects that may lead to geometrical distortions. Depending on the detector used, EBSD measurements over large scale areas can be quite long (typically few hours for the parameters used in this study), which can result in a beam drift over time due to thermal / mechanical changes of the SEM configuration. This effect has been assessed to be negligible in this study by comparing systematically SEM images taken before and after the EBSD map. However, the configuration used for EBSD measurements - which corresponds to a 70$^{\circ}$ tilted specimen with respect to the beam axis - leads to image distortion as can be seen on Fig.~\ref{fig3}b where the Vickers indents are no longer forming a square. Such distortion appears for large scale EBSD maps performed using low magnification, and should be corrected to obtain correct grain shapes.

Several algorithms have been proposed in the literature to deal with 3D EBSD reconstruction \cite{ZHANG2014158,CHARPAGNE2019184,Pirgazi:nb5147}. An in-house procedure is used in this study. In order to correct the geometrical distortion, the displacements of the four corners Vickers indents required to match a square of 1mm size are determined (Fig.~\ref{fig3}c), and the image is then corrected using a bilinear interpolation correction scheme based on these displacements. A typical result is shown on Fig.~\ref{fig3}d where the actual EBSD zone scanned has in fact a trapezoidal shape. All 2D EBSD maps have been corrected according to this procedure. However, this does not ensure that the superposition of the 2D EBSD maps will lead to a 3D EBSD microstructure free of geometrical distortions \cite{PIRGAZI2019223}. The main reason is that slight differences of alignment and / or SEM acquisition parameters between each step are unavoidable since the sample is removed from the SEM for polishing. Therefore, an additional correction procedure is applied to EBSD maps, following the first correction procedure described above. The displacements of the Vickers indents are defined as the ones allowing maximizing the cross-correlation of Euler angles between the current EBSD map and the previous one. In practice, only one Euler angle is used for the cross-correlation, and a Nelder-Mead algorithm is used to perform the maximization. This method is valid as long as two successive EBSD maps are not too different, \textit{i.e.}, when the distance between the two planes is small compared to the grain size, and \textcolor{black}{for materials with no morphological texture} (as for example the method is expected to introduce bias for materials with tilted columnar grains). \textcolor{black}{In addition, Euler angles should be mostly constant in each grain, \textit{i.e.}, without jumps from one voxel to another that can happen due to the FCC symmetry, which has been checked (Appendix~A). In that case, an alternative strategy is to minimize the local misorientation between consecutives maps \cite{PIRGAZI2019223}.} The details about the EBSD corrections are given in Appendix A. Finally, the three Euler angles values are projected on a regular grid - required for the numerical simulations - using nearest value interpolation, allowing subsampling if necessary.
\hspace{-1cm}
\begin{figure}[H]
\centering
\includegraphics[height = 5.0cm]{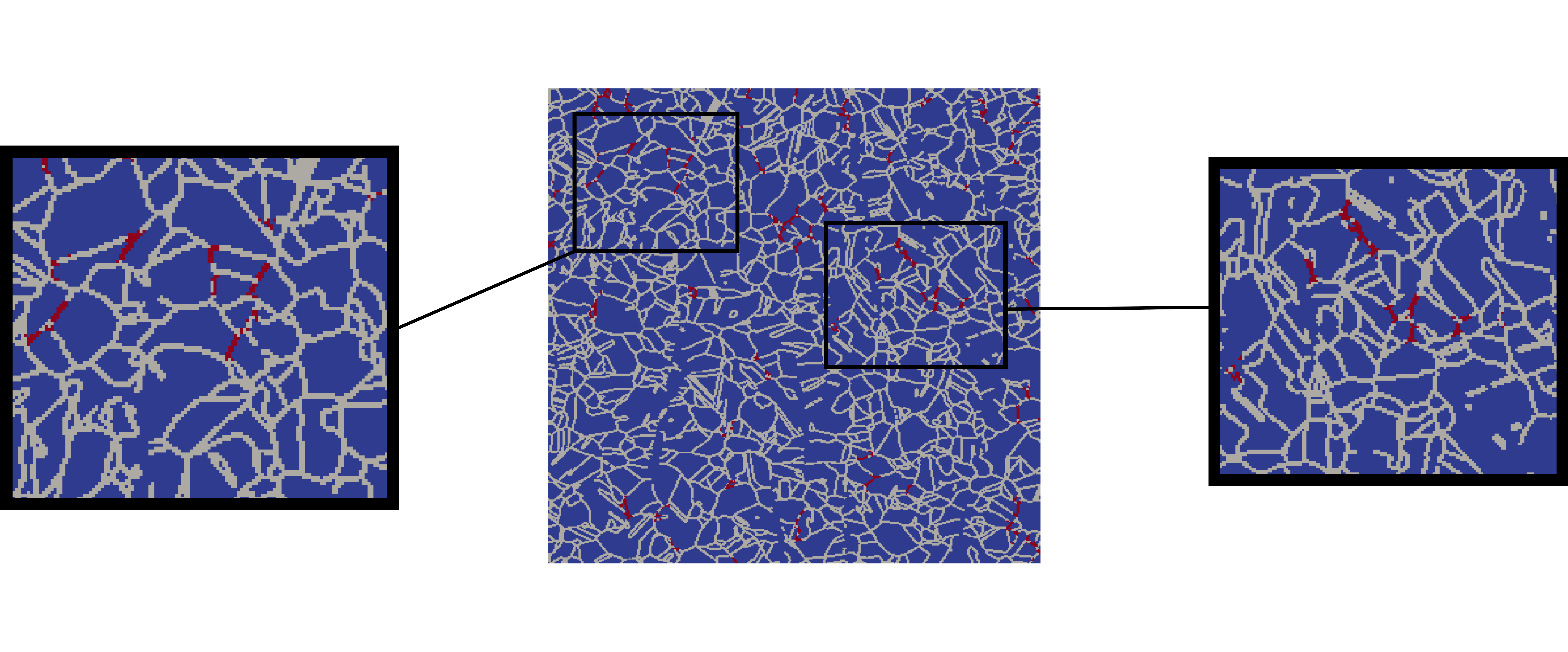}
\vspace{-0.5cm}
\caption{Visualization of the GB (in gray) detected by the \texttt{MTEX} software on the first EBSD map, and projection of the intergranular cracks (in red)}
\label{fig4}
\end{figure}

The 3D EBSD reconstruction is shown on Fig.~\ref{fig3}e, and corresponds to a volume of about $\mathrm{800 \mu m}\ \mathrm{x}\ \mathrm{800 \mu m}\ \mathrm{x}\ \mathrm{20 \mu m}$. The correction procedure is assessed on Fig.~\ref{fig3}f by looking at cross sections of the 3D microstructure, showing \textcolor{black}{a strong reduction of geometrical artefacts for grain shapes compared to the case without corrections (Appendix~A)}. Another artefact reported in the literature \cite{Pirgazi:nb5147} is differences of crystallographic orientations in a given grain on two consecutive 2D EBSD maps due to slight misalignment of the sample inside the SEM. This has been checked in the 3D reconstructed microstructure by looking at in-plane and through thickness Euler angles profiles in several grains. The variations along the thickness - thus between consecutive 2D EBSD maps - are found to be of the same order (few degrees) as the in-plane variations \textcolor{black}{(Appendix~A, Fig.~\ref{figAAA0}b)}, indicating that the problem reported in \cite{Pirgazi:nb5147} is not significant in our study.

\subsubsection{Intergranular cracks}
\label{incra}

The SEM cartography performed at the surface of the sample (Fig.~\ref{fig2}a) allows \textcolor{black}{detection, after automatic binarization and manual corrections, of }intergranular cracks in the frame defined by the Vickers indents. The next step consists of locating these cracks at the surface of the 3D microstructure (Fig.~\ref{fig3}e). First, the GB of the EBSD map are detected using the \texttt{MTEX} software (Fig.~\ref{fig4}) using a misorientation threshold of 10$^{\circ}$. As both the 3D microstructure and the intergranular cracks have been obtained in the same frame, it is possible to project intergranular cracks on the GB, which is shown on Fig.~\ref{fig4}. The spatial resolution of EBSD maps is $\Delta = 2\mu$m, and the maps have been corrected for distortions. Spatial resolution of SEM observations is much higher and distortion free because the sample is perpendicular to the beam axis. Hence, the projection of cracks on GB is not straightforward. For each position corresponding to a crack in the SEM images, the neighbouring pixels are also considered to correspond to a crack up to a distance of \textcolor{black}{$\mathcal{P}\Delta$, to be consistent with the resolution of the EBSD maps. As detailed in Appendix~B, $\mathcal{P} = 1$ is chosen so as to minimize the number of false cracked GB.} The projection is then performed, leading to the results presented in Fig.~\ref{fig4}.

\subsubsection{Grain boundaries 3D characterization}

The \texttt{MTEX} software \cite{bachmann2010} used in the previous section allows \textcolor{black}{only detection of GBs} from a 2D EBSD map, whereas the characterization of the 3D GB is required to compute GB normals. For the analysis of the cracking initiation conditions, at least the computation of GB normals for the free surface GB is needed, which is done with an in-house procedure. The 3D reconstruction of the free-surface GB is briefly described (more details
can be found in Appendix B).

\begin{figure}[H]
\centering
\includegraphics[height = 4.6cm]{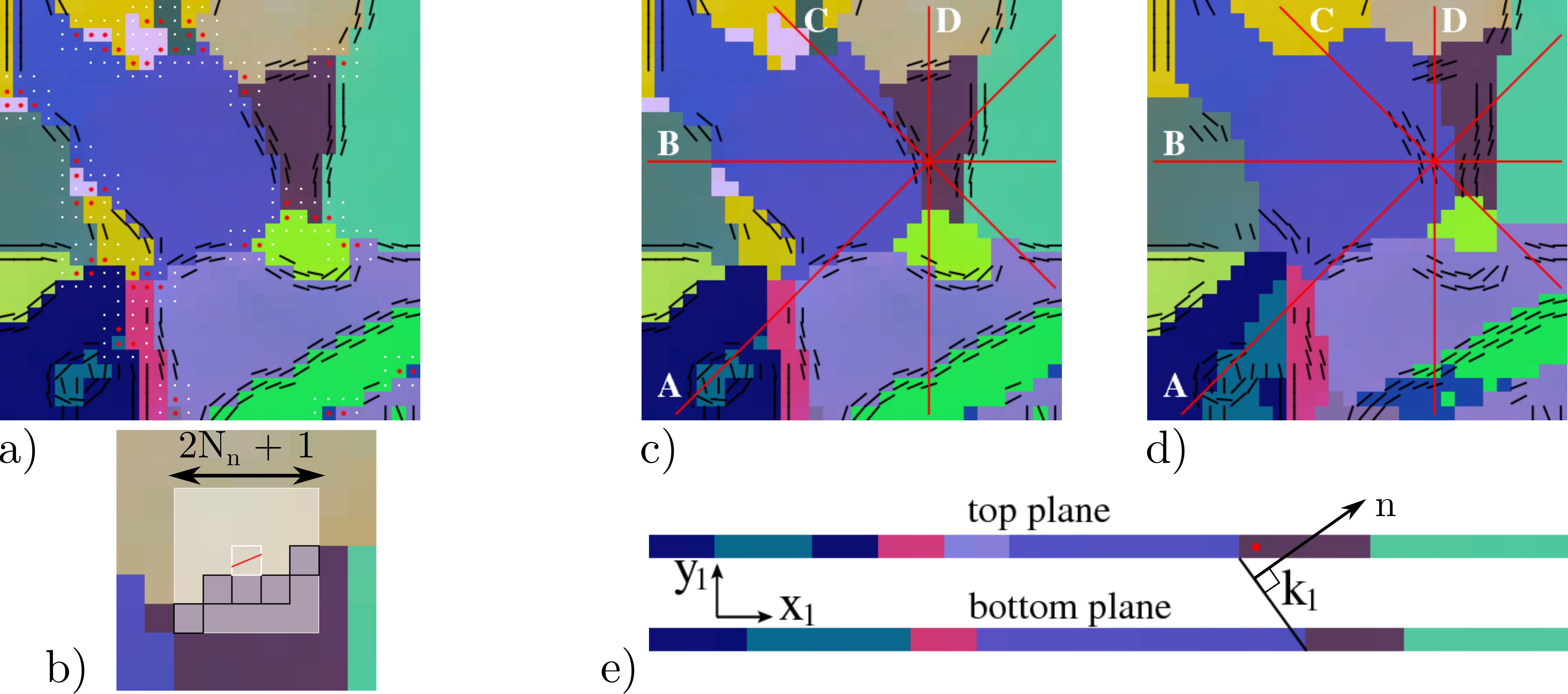}
\caption{(a, b) Detection of GB on the free-surface plane. Black lines denote calculated in-plane GB slopes, red points
  the identified triple points and white points a $3\times 3$
  neighborhood of each triple point where GB detection is avoided (b) Enlarged section of the top plane sketching the calculation
  of the in-plane GB slope (c,d) GB on top and bottom planes: red lines denote vertical planes to detect the out-of-plane GB slope $k_1$ (e).}
\label{fig5}
\end{figure}

The employed method reconstructs
GB from two 2D images stacked atop each other to form a section of a
3D aggregate model of a sample. Each colored voxel forming a regular grid of a 2D image represents a local crystallographic orientation (using the correspondence between the Euler angles and RGB color scheme). On the top horizontal EBSD plane, the method identifies first GB locations  by analyzing local image contrasts, see Fig.~\ref{fig5}a,b, and GB slopes in a selected GB voxel is computed by accounting for nearest-neighbor voxels region of size $(2N_n+1)\times(2N_n+1)$ (thin white square in Fig.~\ref{fig5}b). For each GB voxel identified on a top plane four vertical cross-section planes are then formed and same-colored voxels
identified within each plane in order to calculate the (average)
out-of-plane GB slope, see Fig.~\ref{fig5}c,d,e. Once the in-plane
and out-of-plane GB slopes are known a 3D GB normal $\bm{n}$ is finally calculated and assigned to the corresponding free-surface GB voxel. It has been tested, using 2D images from Voronoi aggregate model with exactly known GB normals, that the accuracy of the above method
improves with the increasing distance between the two considered 2D
image planes. \textcolor{black}{A sensitivity analysis detailed in Appendix~B leads to consider $N_n = 2$ and a distance between the top and bottom planes of $d = 4\mu$m in all computations. It should also be noted that triple points GB (and their close vicinities) are not considered (Fig.~\ref{fig5}) as GB normal computation is not possible. Therefore, intergranular cracks (Fig.~\ref{fig4}) are projected onto the GB locations where GB normals are available.}

\begin{figure}[H]
\centering
\subfigure[]{\includegraphics[height = 4.5cm]{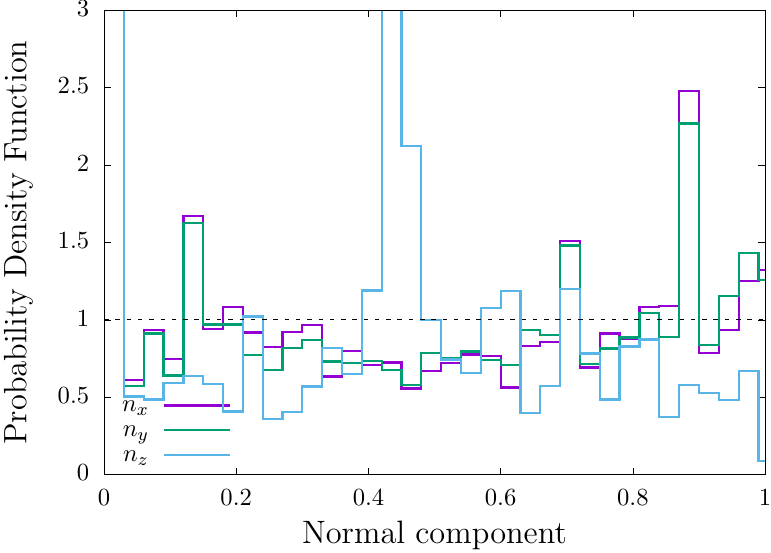}}
\hspace{0.5cm}
\subfigure[]{\includegraphics[height = 4.5cm]{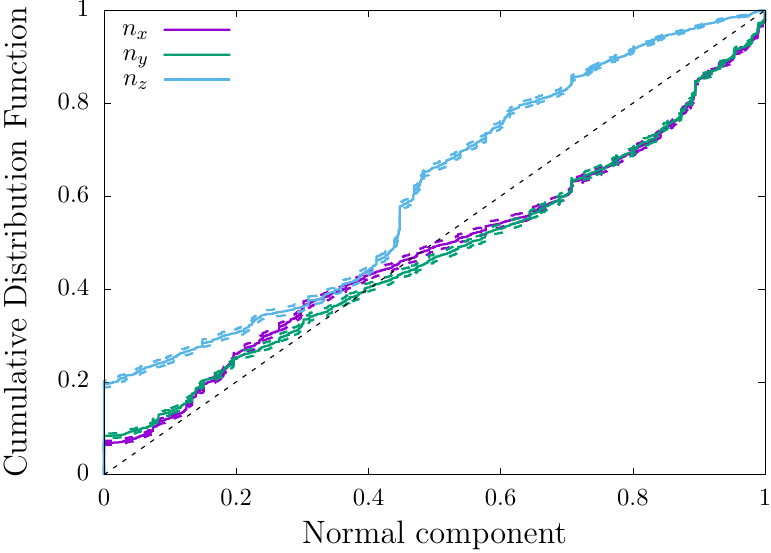}}
\caption{Probability density functions (a) and Cumulative distribution functions (b) of the GB normal $\bm{n}$ components}
\label{fig44}
\end{figure}

\textcolor{black}{The statistical analysis presented hereafter is based on the computation of probability density functions (noted \textit{pdf}) and cumulative distribution functions (\textit{cdf}). For the latter, 95\% confidence bounds are also computed\footnote{\textcolor{black}{as implemented in the \texttt{Matlab} built-in function \textit{ecdf}}}.} \textit{pdf} and \textit{cdf} of the GB $\bm{n}$ normal components are plotted in Fig.~\ref{fig44}a,b. \textcolor{black}{For materials with no morphological texture (grain shapes), the theoretical \textit{pdf}  of the absolute values of the normal components $n_i$ should be equal to 1, and the corresponding \textit{cdf} linear with a slope equal to 1}. In-plane components $n_x$ and $n_y$ distributions are found to be close to these values, \textcolor{black}{showing that the material has no morphological texture. Out-of-plane component $n_z$ exhibits an overrepresentation} of values close to 0 and 0.5, which is probably due to the absence of smoothing for the out-of-plane component as compared to the in-plane components. 

\begin{figure}[H]
\centering
\subfigure[]{\includegraphics[height = 4.5cm]{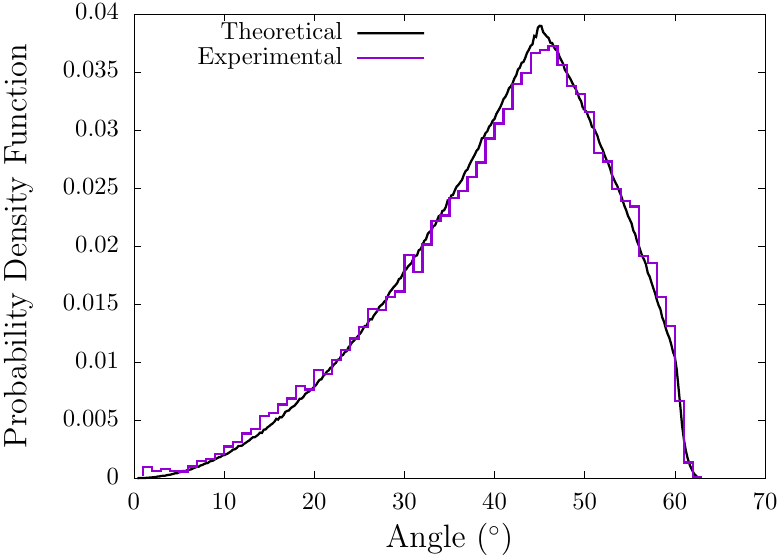}}
\hspace{0.5cm}
\subfigure[]{\includegraphics[height = 4.5cm]{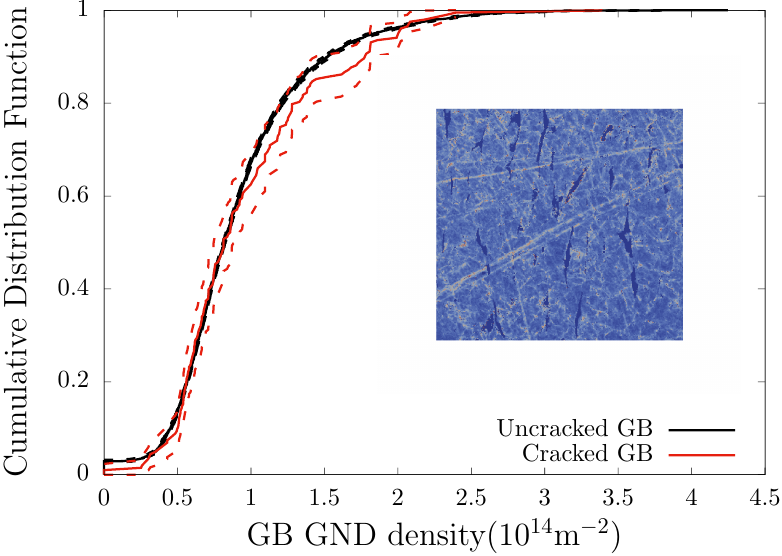}}
\caption{(a) Probability density function of the misorientation angle between random positions on the free-surface. Comparison between the experimental measurements and the theoretical McKenzie distribution for untextured materials (b) Cumulative distribution function of Geometrically Necessary Dislocations (GND) close to uncracked (in black) and cracked (in red) GB. \textcolor{black}{Dotted lines correspond to 95\% confidence bounds.} Inset: GND field at the free surface}
\label{fig444}
\end{figure}

The experimental characterization has allowed \textcolor{black}{reconstruction of} the 3D microstructure - crystallographic orientations and grain shapes - over an extended area, to detect (un-)cracked GB as well as to evaluate GB slopes on the free-surface. The data are then examined to assess local cracking conditions in the next section, keeping in mind some unavoidable imperfections of the experimental characterization for the interpretation of the results. First, the characterization has been performed after the \textcolor{black}{SSRT}, on a deformed material, which could have affected both grain shapes and crystallographic orientations. However, the low level of applied strain ($\sim4\%$) and the mechanical loading condition (uniaxial stress) is expected to lead to minor effects.
For crystallographic orientations, the misorientation angle distribution computed by selecting random positions on the free-surface of the reconstructed microstructure does not show deviations from the theoretical McKenzie distribution for untextured materials (Fig.~\ref{fig444}a). An outcome of having performed the EBSD measurements on the deformed material is that local crystallographic orientations are affected by residual stresses. \texttt{MTEX} software has been used to compute Geometrically Necessary Dislocations (GND) densities associated with orientations gradients at the free surface (Inset Fig.~\ref{fig444}b). As GNDs can be seen as an indicator of local deformation incompatibilities, average values close to GB\footnote{The averaging is performed using the first neighbors, thus on a disk of 2 $\mu\mathrm{m}$ radius.} have been computed, and cumulative distribution functions for uncracked and cracked GB are shown on Fig.~\ref{fig444}b. No significant difference is observed, thus cracking can not be associated with GNDs in this study. However, as GNDs computations depend strongly on the accuracy of crystallographic orientations and GND may arise at a distance of GB smaller than the spatial resolution of the EBSD maps, additional measurements are required.

\subsection{Analysis of experimental data}
\label{anaexpdata}
\subsubsection{GB normals}

\begin{figure}[H]
\centering
\subfigure[]{\includegraphics[height = 4.5cm]{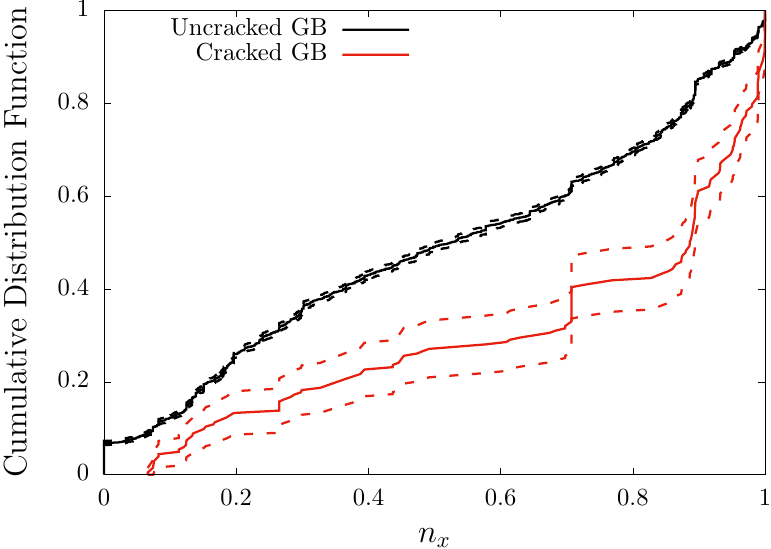}}
\hspace{0.5cm}
\subfigure[]{\includegraphics[height = 4.5cm]{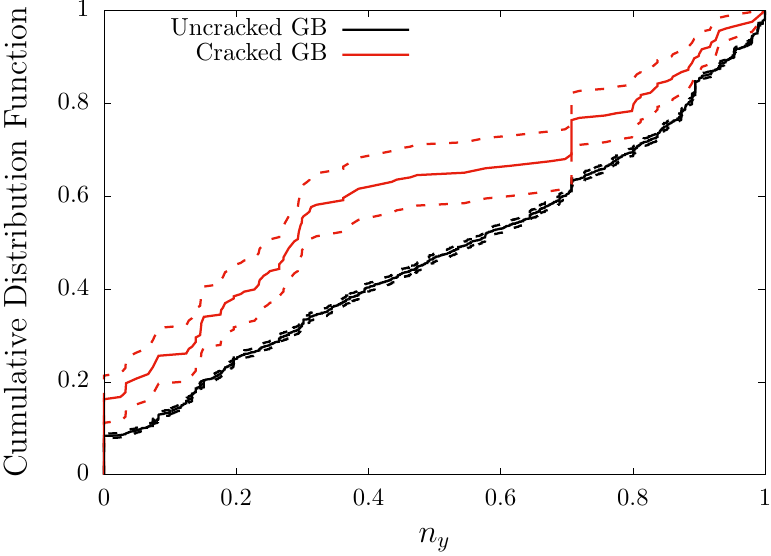}}
\caption{Cumulative distribution functions of the GB normal $\bm{n}$ components for uncracked (in black) and cracked (in red) GB (a) $n_x$ and (b) $n_y$. \textcolor{black}{Dotted lines correspond to 95\% confidence bounds.}}
\label{fig55}
\end{figure}

The \textit{cdf} of GB normals are first assessed in this section. Fig.~\ref{fig55} shows the \textit{cdf} of the normal component $n_x$, the $x-$axis corresponding to the loading axis during the \textcolor{black}{SSRT}. As already shown in Fig.~\ref{fig44}b, the \textit{cdf} of uncracked GB (which correspond to the majority of the GB) \textcolor{black}{almost} follows a linear relationship of slope 1 with the absolute value of the normal component, as \textcolor{black}{expected} for materials \textcolor{black}{with no morphological texture}. Cracked GB exhibit a significant deviation from uncracked GB, where cracked GB have statistically higher values of $n_x$. Accordingly, the \textit{cdf} of the two other normal components - $n_y$ (Fig.~\ref{fig55}b) and $n_z$ - of cracked GB are significantly shifted towards the lower values as compared to uncracked GB.

These results, consistent with previous studies based only on 2D observations (see, \textit{e.g.}, \cite{west}), support the dependence of IGSCC of irradiated stainless steels to the local stress conditions, as the normal stress acting on a GB is $\sigma_{nn} = \sigma\, n_x^2$ (assuming an uniaxial stress state of magnitude $\sigma$). Of course, the local stress conditions may deviate from uniaxial stress conditions due to grain-grain interactions, but higher intergranular normal stresses are in majority related to well-oriented boundaries, thus with high values of $|n_x|$ for uniaxial loading conditions \cite{DIARD200273}. Interestingly, Fig.~\ref{fig55} shows that a significant number of cracked GB are not well-oriented with respect to the mechanical loading axis, and thus that the GB orientation is not a sufficient condition for cracking.

\subsubsection{Slip transmission criteria}

Experimental observations based on slip traces at the surface of irradiated austenitic stainless steels samples tested in PWR environment have been reported to show a correlation between slip discontinuity and cracking \cite{MCMURTREY2015305}. \textcolor{black}{Several slip transmission criteria have been proposed in the literature to predict such discontinuities \cite{slipt}. The use of a reliable slip transmission criterion is thus a potential way to predict GB that are most susceptible to cracking. However, assessment of these criteria with respect to experimental observations remains scarce \cite{ALIZADEH2020408}, and, to the authors' knowledge, no such assessment has been done yet for irradiated austenitic stainless steels.} The experimental data obtained in this study \textcolor{black}{do not allow to assess the reliability of slip transmission criteria, as the (dis-)continuity of slip traces at GB has not been examined systematically. However,} as GB have been fully characterized with crystallographic orientations on each side and GB normal orientation, \textcolor{black}{correlations between slip transmission criteria and cracking are assessed.} Considering a GB of normal $\bm{n}_{\Gamma}$ between two grains $A$ and $B$, each of them having a set of slip systems of slip direction $\bm{d}^{A,B}_i$ and slip plane normal $\bm{n}^{A,B}_i$, slip transmission parameters $N_{ij}^{A,B}$ can be written as:
\begin{equation}
  N_{ij}^{A,B} = \mathcal{F}\left(\bm{n}_{\Gamma},\bm{d}^{A,B}_{i,j},\bm{n}^{A,B}_{i,j}    \right) 
\end{equation}
The slip transmission criteria used in this study are summarized in Tab.~\ref{tab1}. More complicated slip transmission criteria exist \cite{slipt}, but depend on additional material parameters that need to be calibrated and are thus not used. Slip transmission is evaluated based on the 12 x 12 matrix $N_{ij}^{A,B}$ coefficients defined for each GB. In the following, the maximal value $N_{max}^{A,B} = max_{i,j} |N_{ij}^{A,B}|$ is used as an indicator for slip transmission. In addition, Schmid factors are also computed as low values have been correlated with intergranular cracking \cite{MCMURTREY20113730}.

\begin{table}[H]
  \centering
  \begin{tabular}{c||c}
    & Parameter  $N_{ij}^{A,B}$ \\
    \hline
    \hline
    Livingston-Chalmers & $(\bm{n}_i^A \cdot \bm{n}_j^B )(\bm{d}_i^A \cdot \bm{d}_j^B ) + (\bm{n}_i^A \cdot \bm{d}_j^B )(\bm{n}_j^B \cdot \bm{d}_i^A )$      \\
    \hline
    Luster-Morris & $(\bm{n}_i^A \cdot \bm{n}_j^B )(\bm{d}_i^A \cdot \bm{d}_j^B )$     \\
    \hline
    Shen-Wagoner-Clark & $(\bm{l}_i^A \cdot \bm{l}_j^B )(\bm{d}_i^A \cdot \bm{d}_j^B )$ with $\bm{l}_i^{A,B} = (\bm{n}_i^{A,B} \times \bm{n}_{\Gamma}) / |\bm{n}_i^{A,B} \times \bm{n}_{\Gamma}|$   \\
    \hline
    Lee-Robertson-Birnbaum &  $(\bm{l}_i^A \cdot \bm{l}_j^B)$ with $\bm{l}_i^{A,B} = (\bm{n}_i^{A,B} \times \bm{n}_{\Gamma}) / |\bm{n}_i^{A,B} \times \bm{n}_{\Gamma}|$   \\       
  \end{tabular}
  \caption{Summary of the slip transmission parameters used in this study \cite{slipt}}
  \label{tab1}
\end{table}

\begin{figure}[H]
\centering
\subfigure[]{\includegraphics[height = 4.5cm]{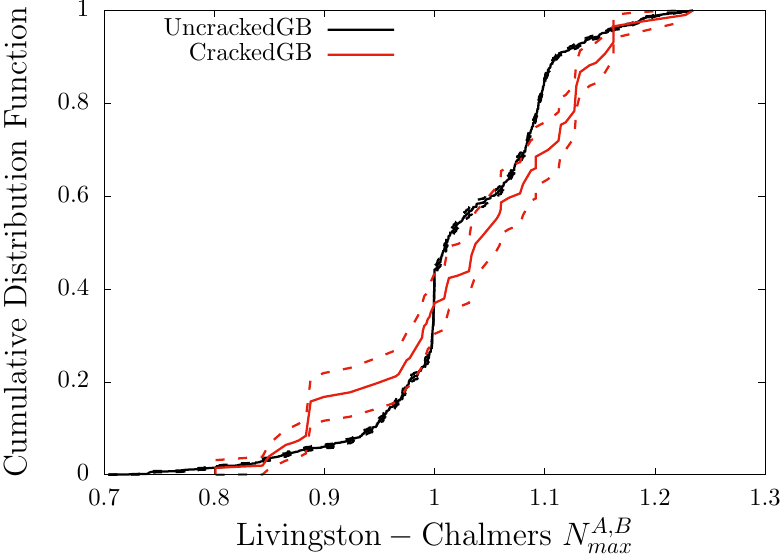}}
\hspace{0.5cm}
\subfigure[]{\includegraphics[height = 4.5cm]{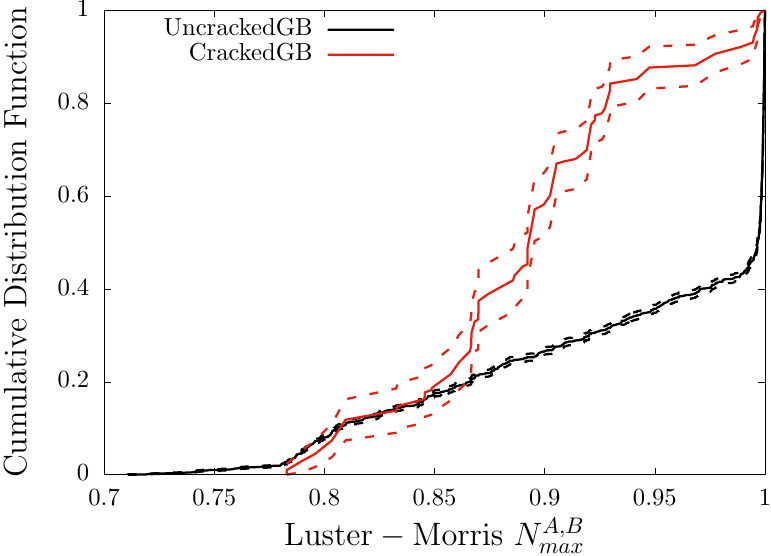}}
\subfigure[]{\includegraphics[height = 4.5cm]{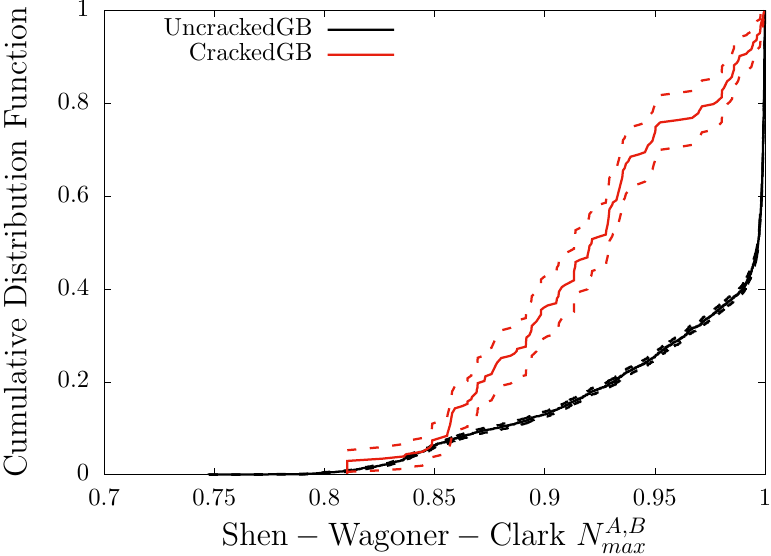}}
\hspace{0.5cm}
\subfigure[]{\includegraphics[height = 4.5cm]{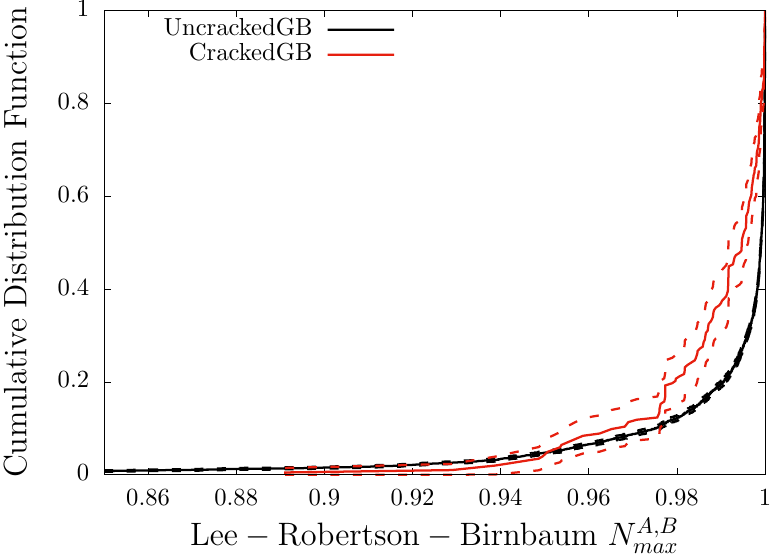}}
\caption{\textcolor{black}{Cumulative distribution functions of slip transmission parameter $N_{max}^{A,B}$ for uncracked (in black) and cracked (in red) GB, using (a) Livingston-Chalmers (b) Luster-Morris (c) Shen-Wagoner-Clark and (d) Lee-Robertson-Birnbaum models. Dotted lines correspond to 95\% confidence bounds.}}
\label{fig6}
\end{figure}

The cumulative distribution functions of $N_{max}^{A,B}$ are computed for both uncracked and cracked GB, and the results are shown on Fig.~\ref{fig6}. \textcolor{black}{For Livingston-Chalmers model (Fig.~\ref{fig6}a)}, no significant difference between cracked and uncracked GB is observed, whereas a slight difference appears \textcolor{black}{for Lee-Robertson-Birnbaum model} (Fig.~\ref{fig6}d). Significant differences are found for the Luster-Morris and Shen-Wagoner-Clark models (Fig.~\ref{fig6}b,c), with lower values of $N_{max}^{A,B}$ for cracked GB. Luster-Morris and Shen-Wagoner-Clark models differ only by the fact that the first one used the slip plane normals, while in the second the cross-product of the slip plane normal by the GB normal. The differences between cracked and uncracked GB for both models are rather similar, supporting the fact that GB normals may not be a key ingredient \textcolor{black}{of the correlation with cracking. Thus, only Luster-Morris model is considered in the following}. 

\subsubsection{Intergranular cracking correlations}

The previous sections have shown correlations between intergranular cracking and both GB normal well-oriented with respect to the loading axis and low values of Luster-Morris parameters. These two criteria may be related to higher local intergranular normal stresses, which thus appear to play a major role for IGSCC of irradiated austenitic stainless steels in PWR environment. A correlation using both GB normal and Luster-Morris parameter is assessed in the following, defined as:
\begin{equation}
  \left\{
  \begin{array}{l}
    |n_x| \geq \alpha_n \\
    N_{max}^{\mathrm{Luster-Morris}} \leq \alpha_N 
  \end{array}  
  \right.
  \label{eqcrack}
\end{equation}

\begin{figure}[H]
\centering
\subfigure[]{\includegraphics[height = 3.7cm]{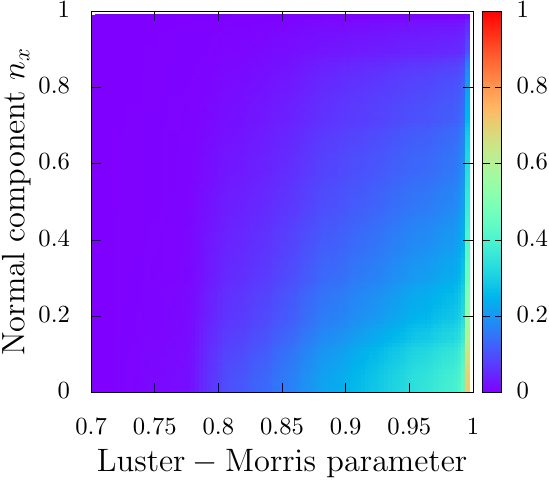}}
\subfigure[]{\includegraphics[height = 3.7cm]{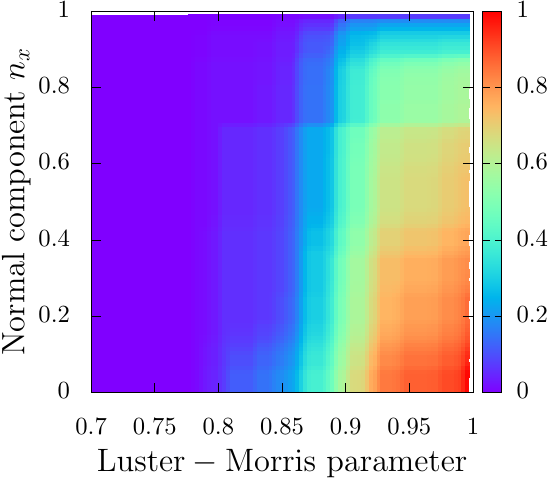}}
\subfigure[]{\includegraphics[height = 3.7cm]{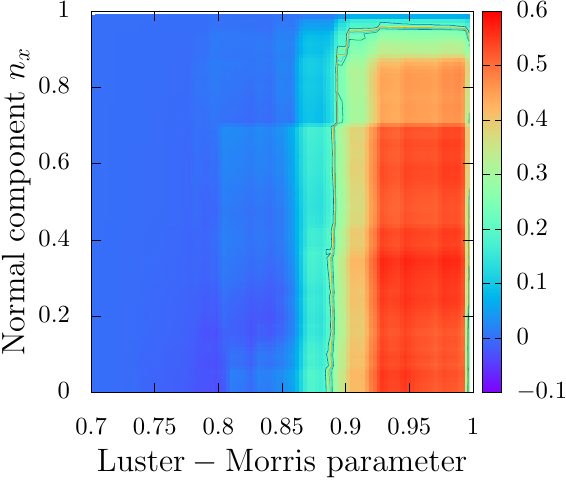}}
\caption{Percentage of GB fulfilling Eq.~\ref{eqcrack} as a function of the parameters $\{\alpha_n;\alpha_N\}$ for (a) uncracked and (b) cracked GB. (c) Differences between (a) and (b)}
\label{fig7}
\end{figure}

The percentage of GB that fulfil Eq.~\ref{eqcrack} as a function of the parameters $\{\alpha_n;\alpha_N\}$ are plotted on Fig.~\ref{fig7}a for uncracked GB for reference to be compared to the case of cracked GB in Fig.~\ref{fig7}b. A significant difference is observed between the two distributions, which can be highlighted by looking at the difference between them, which is done on Fig.~\ref{fig7}c. It is found that, \textcolor{black}{for $\{\alpha_n;\alpha_N\} = \{0.6;0.95\}$, Eq.~\ref{eqcrack} is fulfilled by about 65\% of the cracked GB, while only by 15\% of the uncracked GB}. Interestingly, the threshold value for the Luster-Morris parameter is similair to the one used in \cite{HAOUALA2020102600} and based on experimental observations of slip transmission. \textcolor{black}{These results make clear that local mechanical state at GB is a factor influencing IGSCC of irradiated austenitic stainless in PWR environment, but is definitely not a sufficient condition for cracking as 35\% of the cracked GBs do not fulfil Eq.~\ref{eqcrack}. This has already been pointed out in early studies \cite{WEST2011142}  and will be discussed in Section~\ref{sec4}}.

\begin{figure}[H]
\centering
\subfigure[]{\includegraphics[height = 4.5cm]{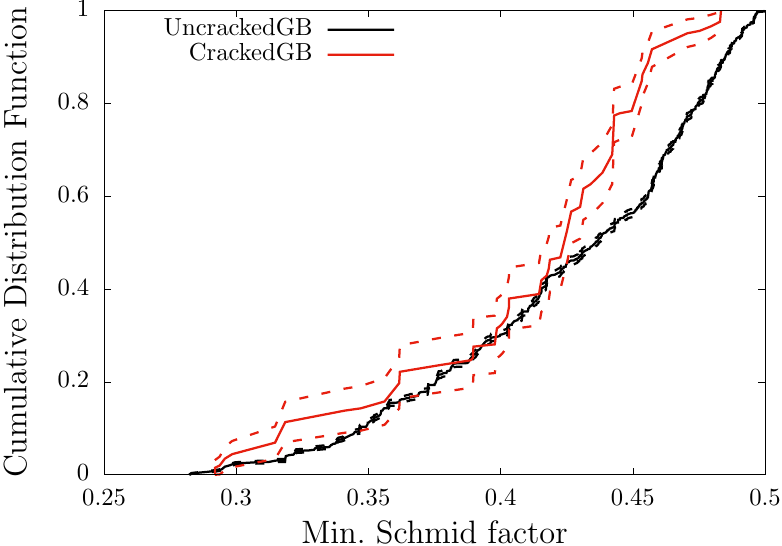}}
\hspace{0.5cm}
\subfigure[]{\includegraphics[height = 4.5cm]{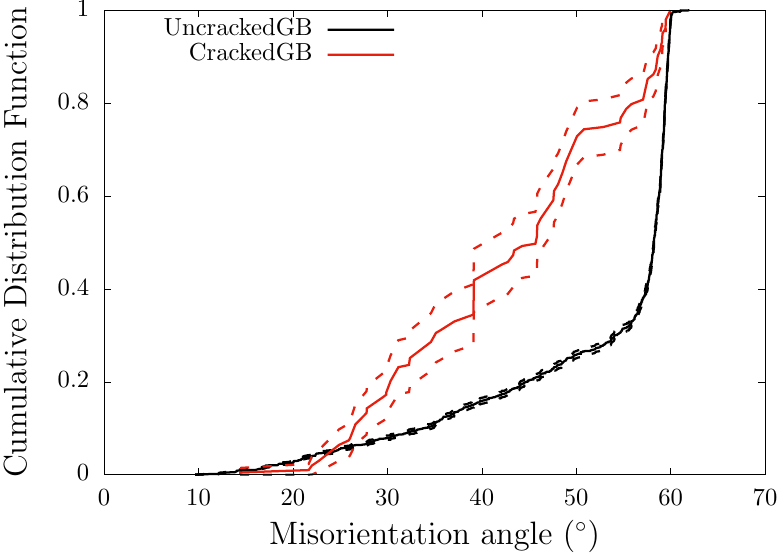}}
\caption{\textcolor{black}{Cumulative distribution functions of (a) minimal Schmid factor (b) misorientation angle for uncracked (in black) and cracked (in red) GB. Dotted lines correspond to 95\% confidence bounds.}}
\label{fig60}
\end{figure}

\textcolor{black}{In order to understand in more details the physical origin of the correlation observed between cracking and low values of the Luster-Morris parameter, additional correlations are assessed.} As previous studies have found correlations between intergranular cracking and low values of Schmid factor on one side of the GB \cite{WEST2011142} \textcolor{black}{based on the argument that such situation leads to higher stresses}, the \textit{cdf} of minimal Schmid factor (over the two grains forming the GB) has been computed and shown in Fig.~\ref{fig60}a. The minimal Schmid factor for cracked GB appears to be also lower for these experimental data, although the difference is weak. More importantly, misorientation angle between grains defining GBs is computed and distributions are shown on Fig.~\ref{fig60}b. For uncracked GB, about 60\% have a misorientation angle close to $60^{\circ}$, which corresponds to $\Sigma_3$ twin boundary. For cracked GB, this proportion is considerably lower, about 20\%. This result is consistent with previous studies showing that special GB, and especially $\Sigma_3$, are less susceptible to IGSCC in austenitic stainless steels \cite{MCMURTREY20113730,GERTSMAN20011589}. 

\begin{figure}[H]
\centering
\subfigure[]{\includegraphics[height = 4.5cm]{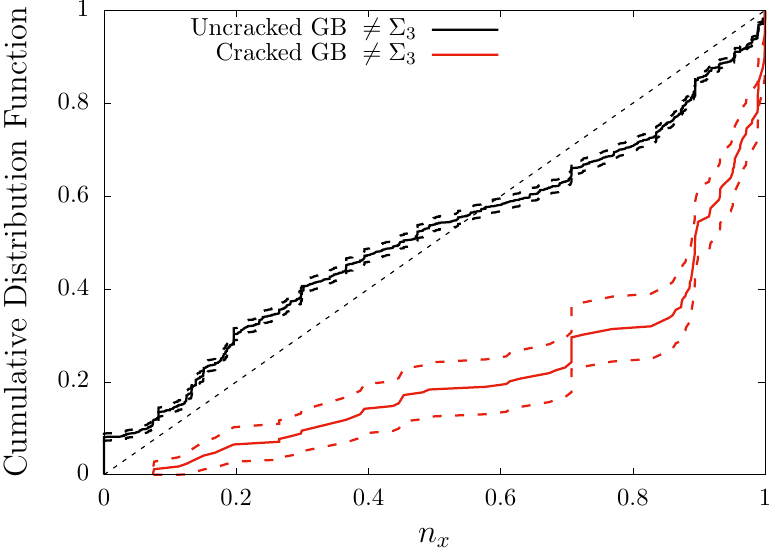}}
\hspace{0.5cm}
\subfigure[]{\includegraphics[height = 4.5cm]{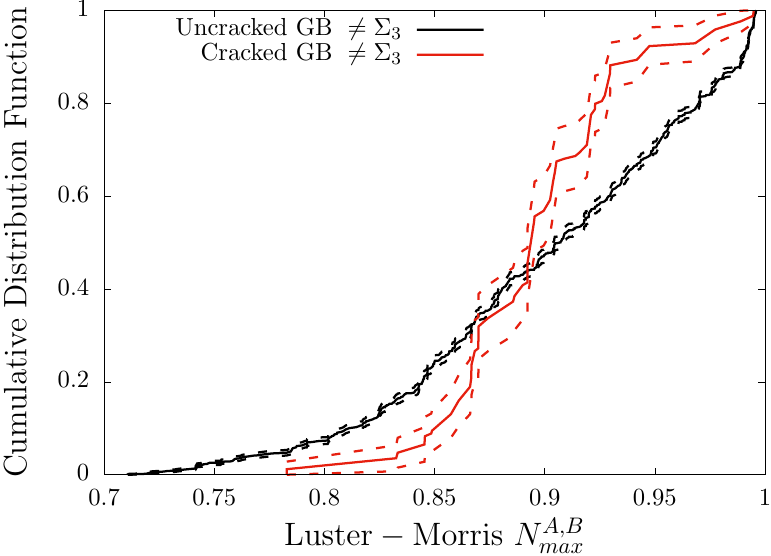}}
\caption{\textcolor{black}{Cumulative distribution functions of (a) normal component $n_x$ and (b) slip transmission parameter $N_{max}^{A,B}$ using Luster-Morris model, for uncracked (in black) and cracked (in red) non $\Sigma_3$ GB. Dotted lines correspond to 95\% confidence bounds.}}
\label{figsigma3}
\end{figure}

Based on this result, correlations between cracking and well-oriented GB (Figs.~\ref{fig55}a) / low values of the Luster-Morris parameter (Fig.~\ref{fig6}b) are reassessed by considering only non $\Sigma_3$ GB, assuming that $\Sigma_3$ GB are less susceptible to cracking. Distributions of GB normal component $n_x$ (Fig.~\ref{figsigma3}a) are similar to the ones obtained considering all GB (Fig.~\ref{fig55}a). The distribution of Luster-Morris parameter for uncracked GB (Fig.~\ref{figsigma3}b) is completely different to the one shown in Fig.~\ref{fig6}b, while the distribution for cracked GB is weakly affected. This is due to the fact that Luster-Morris parameter $N_{max}^{A,B} \simeq 1$ for $\Sigma_3$ GB.  However, cracked GB still exhibit lower values.

\textcolor{black}{The results shown in Fig.~\ref{figsigma3}b indicate that the correlation between cracking and low values of the Luster-Morris parameter (Fig.~\ref{fig6}b) has (at least) two different origins. The first one comes from the overrepresentation of $\Sigma_3$ type GB on uncracked GB compared to cracked GB. Restricting to non $\Sigma_3$ GB, significant statistical differences are still observed on Fig.~\ref{figsigma3}b, that are attributed to the effect of slip discontinuity at GBs on cracking. As stated before, additional studies are still required to assess if Luster-Morris criterion allows effectively to predict slip (dis)-continuity at GBs in irradiated austenitic steels. Note that such assessment has been done for Aluminum \cite{ALIZADEH2020408}, leading to good agreement with experimental observations associated with some additional criteria. Nevertheless, the experimental results obtained in this study confirm that IGSCC cracking of irradiated austenitic steels occurs preferentially on well oriented non $\Sigma_3$ GB, as already shown in previous studies, but also on GB that have low values of Luster-Morris parameter (Fig.~\ref{figsigma3}b), which is a new observation to the authors'knowledge.}


\section{Numerical assessment}
\label{sec3}

The micromechanical analysis performed in the previous section has shown that intergranular cracking of irradiated austenitic stainless in PWR environment is related to the local intergranular (normal) stresses, although neither sufficient nor necessary for cracking. In this section, numerical simulations based on the 3D microstructure obtained in the experimental part are performed to assess the strengths and weaknesses of such micromechanical modelling approach, as well as to provide directions for further studies. 

\subsection{Constitutive equations}

Physically based Crystal Plasticity (CP) constitutive equations are widely used to describe plasticity of single crystals, accounting for the anisotropy induced by the slip systems and hardening mechanisms \cite{ROTERS20101152}. Simulations on polycrystalline aggregates using CP equations allow to estimate the homogenized behavior, or to evaluate intra- and inter-granular stresses, as for example done in \cite{GONZALEZ201449,HURE2016231}. Dedicated physically-based crystal plasticity constitutive equations have been developed for irradiated materials, accounting for the presence of irradiation defects \cite{doi:10.1080/14786435.2011.634855,BARTON2013341,ERINOSHO2015170}. The key ingredients of such models are the influence of irradiation defects on hardening as well as the removal of irradiation defects by dislocations. Few constitutive equations have been proposed for irradiated austenitic stainless steels irradiated in PWR conditions \cite{HURE2016231,MONNET2019316}. After calibration, these models reproduce the evolution of the polycrystalline behavior with irradiation. In the following, the model described in \cite{HURE2016231} is used. The three main sets of equations of the model are described below. For each slip system $\alpha$ corresponding to a slip direction $\bm{m}^{\alpha}$ and a slip plane of normal $\bm{n}^{\alpha}$ (defining a Schmid tensor $\bm{M}^{\alpha} = \bm{m}^{\alpha} \otimes \bm{n}^{\alpha}$), the evolution of shear strain $\dot{\gm}^\al$ is given by:
\be
  \dot{\gm}^\al=\left\langle\frac{|\tau^\al| - \tau_c^\al}{K_0}\right\rangle^n {\rm sign}(\tau^\al),
  \quad\hbox{with}\quad\langle x\rangle=\left\{ \begin{array}{ll}
	x &; x>0\\
        0 &; x\le 0
	\end{array}\right.
  \label{eq_gm}
\ee
where $\tau^\al = \bm{\sigma} : \bm{M}^{\alpha}$ and $\tau_c^\al$ are the resolved shear stress and the Critical Resolved Shear Stress (CRSS), respectively, and $\bm{\sigma}$ the Cauchy stress tensor. For FCC materials, 12 slip systems $<111>(110)$ are considered. The viscoplastic regularization through the parameters $K_0$ and $n$ is used for numerical reasons, and chosen such as to have a nearly time-independent response. The evolution of the CRSS is given by:
\be
  \tau_c^\al = 
  \tau_0 +
  \tau_a \exp{\left(-\frac{| \gamma^{\alpha}  |}{\gamma_0}     \right)} +
  \mu\sqrt{\sum\limits_{\bt=1}^{12} a^{\al\bt}r_D^\bt}+\mu\al_L\sqrt{\sum\limits_{p=1}^4 r_L^p}
  \label{eq_tc}
  \ee
  where $\tau_0$ is an effective lattice friction stress that may account for Hall-Petch effects when calibrated with respect to polycrystalline aggregates simulations. A so-called avalanche term introducing a softening after yielding is set by the parameters $\tau_a$ and $\gamma_0$. $r_D^\al$ is a normalized dislocation density in slip system $\al$ (normalization factor $b^2_D$, with Burgers vector $b_D = 2.54\ 10^{-10}$m), $\mu$ and $a^{\al\bt}$ are respectively the macroscopic shear modulus and $12\times 12$ matrix (with 6 independent parameters) of long-range interactions between dislocations. $r_L^p$ is a normalized Frank loop density in slip plane $p$ (normalization factor $b^2_L \phi_L$, with Burgers vector $b_L = 2.08.10^{-10}$m), $\phi_L$ the mean size of Frank loops that depends on irradiation level and $\al_L$ sets the relative contribution of Frank loops to hardening. The evolutions of dislocation density and Frank loops density are:  
\be
\begin{aligned}
  \dot{r}_D^\al &=
  \left(\frac{1}{\kp}\sqrt{\sum\limits_{\bt=1}^{12} b^{\al\bt}r_D^\bt}+
  \frac{1}{\kp}\sqrt{K_{dl}\sum\limits_{p=1}^4 r_L^p}-G_c r_D^\al\right)|\dot{\gm}^\al| \\
  \dot{r}_L^p &= -A_L (r_L^p-r_L^{sat})\left(\sum_{\al\in {\rm plane}\ p}^3\!\!\!\! r_D^\al\right) 
  \left(\sum_{\al\in {\rm plane}\ p}^3\!\!\!\! |\dot{\gm}^\al|\right) \\
\end{aligned}
  \label{eq_rl}
\ee
where $b^{\al\bt}$ is a matrix of interactions between dislocations,
being of the same shape as $a^{\al\bt}$. Parameters $\kp$ and $G_c$ set the multiplication and annihilation mechanisms, respectively. The
irradiation effects are modelled by adding a term to the multiplication
part, with $K_{dl}$ being a coefficient of effective interaction
between dislocations and Frank loops. The evolution of Frank loops depends on the parameter $A_L$  which is the annihilation dimensionless area (rescaling factor $b_L^3/\phi_L$) of Frank loops and $r_L^{sat}$ is
a stabilized value of normalized defect density which depends on the
irradiation dose. Since \textcolor{black}{removal} of Frank loops by mobile dislocations
occurs only within the plane of the loop, only slipping in this plane
can contribute to the evolution of defect density ($\al\in {\rm
plane}\ p$). Anisotropic elasticity is finally considered, with non-zero parameters of the elastic fourth order tensor $C_{11}=C_{22}=C_{33}$,
$C_{12}=C_{13}=C_{23}$ and $C_{44}=C_{55}=C_{66}$ in Voigt notations.

Slip transmission at GB has been shown to be a key ingredient in IGSCC of irradiated austenitic stainless steels. \textcolor{black}{The data obtained in this study supports this statement through the correlation observed between low values of the Luster-Morris parameter - associated with less slip transmission - and cracking (Fig.\ref{figsigma3}b).} Using the constitutive equations detailed above in polycrystalline aggregates simulations without any additional modelling corresponds to the hypothesis of fully-transparent GB. Intergranular stresses will result only from grain-grain interactions, which is clearly a limitation. The model proposed recently in \cite{HAOUALA2020102600} considers a simple modification of the evolution law for dislocation density as:
\be
  \dot{r}_D^\al =
  \left(\max{ \left[\frac{1}{\kp}\sqrt{\sum\limits_{\bt=1}^{12} b^{\al\bt}r_D^\bt}, \frac{K_s^{\alpha}}{d_b^{\al}}\right]}+
  \frac{1}{\kp}\sqrt{K_{dl}\sum\limits_{p=1}^4 r_L^p}-G_c r_D^\al\right)|\dot{\gm}^\al| 
  \label{eq_rd2}
\ee
where $K_s^{\alpha}$ is an additional parameter and $d_b^{\al}$ is the normalized (with $b_D$) minimal distance to a GB along slip system $\alpha$. The parameter $K_s^{\alpha}$ depends on the value of the Luster-Morris parameter such as:
\be
   K_s^{\alpha} =\left\{ \begin{array}{ll}
	K_s^0 & \ \mathrm{if}\ \forall \beta\ N_{\alpha,\beta} < \alpha_N\\
        0 & \mathrm{otherwise}
	\end{array}\right.
  \label{Ks}
  \ee
  where $\beta$ corresponds to the slip systems of the closest grain (at a distance $d_b^{\al}$). Far from the GB, Eq.~\ref{eq_rd2} and~\ref{Ks} tend to Eq.~\ref{eq_rl}, \textit{i.e.} no effect of GB on the local mechanical behavior. Close to the GB, a strong increase of the density of dislocations is obtained in slip systems where slip transmission is unlikely, leading to a strong hardening through Eq.~\ref{eq_tc}. The parameters of the constitutive equations are taken from \cite{HURE2016231} for a dose of 2dpa, consistent with the average dose of the irradiated layer of the sample (Fig.~\ref{fig1}a). The parameter $K_s^0$ was taken to $K_s^0 = 5$ in \cite{HAOUALA2020102600}. \textcolor{black}{$K_s^0 = 0$ is also used to recover Eq.~\ref{eq_rl}}. From Eq.~\ref{eq_rd2}, it can be noticed that the modelling of GB has an effect for:
  \begin{equation}
     d_b^{\al} \leq K_s^0  \, \kappa \, \left(\sqrt{\sum\limits_{\bt=1}^{12} b^{\al\bt}r_D^\bt}\right)^{-1}
    \end{equation}
thus the discretization used in simulations should be finer to have any effects. All the parameters are summarized in Tab.~\ref{tab2}.

\begin{table}[H]
  \centering
  \scalebox{0.85}{
    \centering
    \begin{tabular}{cccccccccc}
      \hline
      \hline
      $C_{11}$ & $C_{12}$ & $C_{44}$ & $n$ & $K_0$ & $\tau_0$ & $\mu$ & $\kappa$ & $G_c$ & $r_D^0$  \\
      \hline
      (GPa)  &  (GPa)  & (GPa)  & & (MPa) & (MPa) & (GPa)  &    &   & \\
      \hline
      \hline
 199     &   136      &  105      &     15   & 10  & 88   & 65.6    & 42.8  & 10.4 & $3.66\ 10^{-11}$\\
      \hline
      \hline
      $\alpha_L$ & $K_{dl}$ & $A_L$ & $r_L^{sat}$ & $r_L^0$ & $\tau_a$ & $\gamma_0$ &  $K_s^0 $\\
      \hline
      &         &       &           &    & (MPa)   &    &  \\
      \hline
      \hline
  0.44    &   $0.25\ 10^{-6}$      &  $4.48\ 10^{8}$     &  $3.78\ 10^{-6}$          &  $4.72\ 10^{-6}$  &  50  &   $5\ 10^{-3}$  &  \textcolor{black}{$[0,5]$} \\
      \hline
      \hline
    \end{tabular}
    }
  \caption{Parameters of the crystal plasticity constitutive equations}
  \label{tab2}
  \end{table}

The constitutive equations have been implemented in the \texttt{MFront} code generator \cite{Mfront} under finite strain settings using a fully implicit integration scheme solved by a Newton-Raphson algorithm. Details about the numerical implementation can be found in \cite{HURE2016231}.
  
\subsection{FFT simulations}

\begin{figure}[H]
\centering
\subfigure[]{\includegraphics[height = 6.0cm]{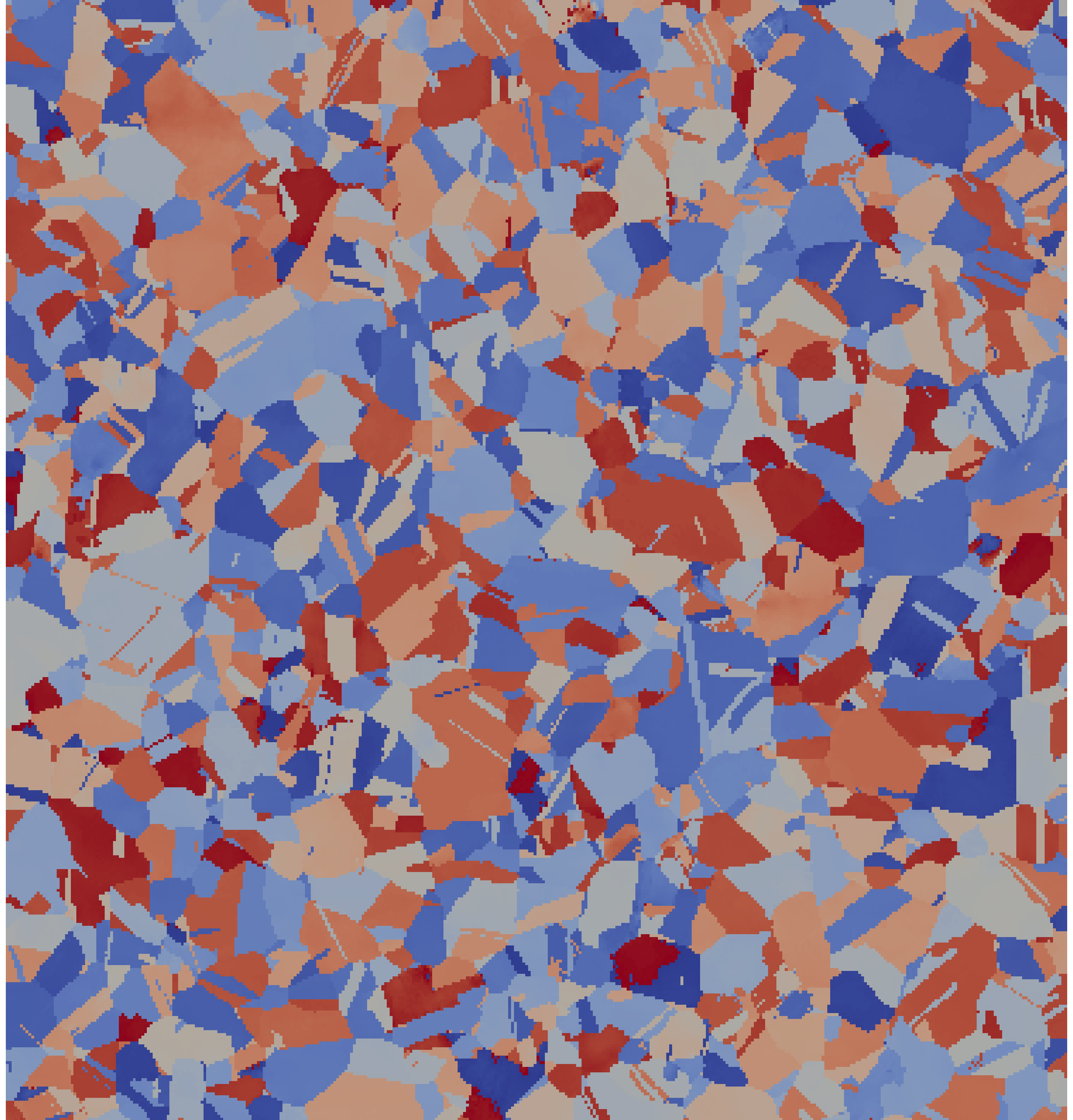}}
\hspace{.9cm}
\subfigure[]{\includegraphics[height = 6.0cm]{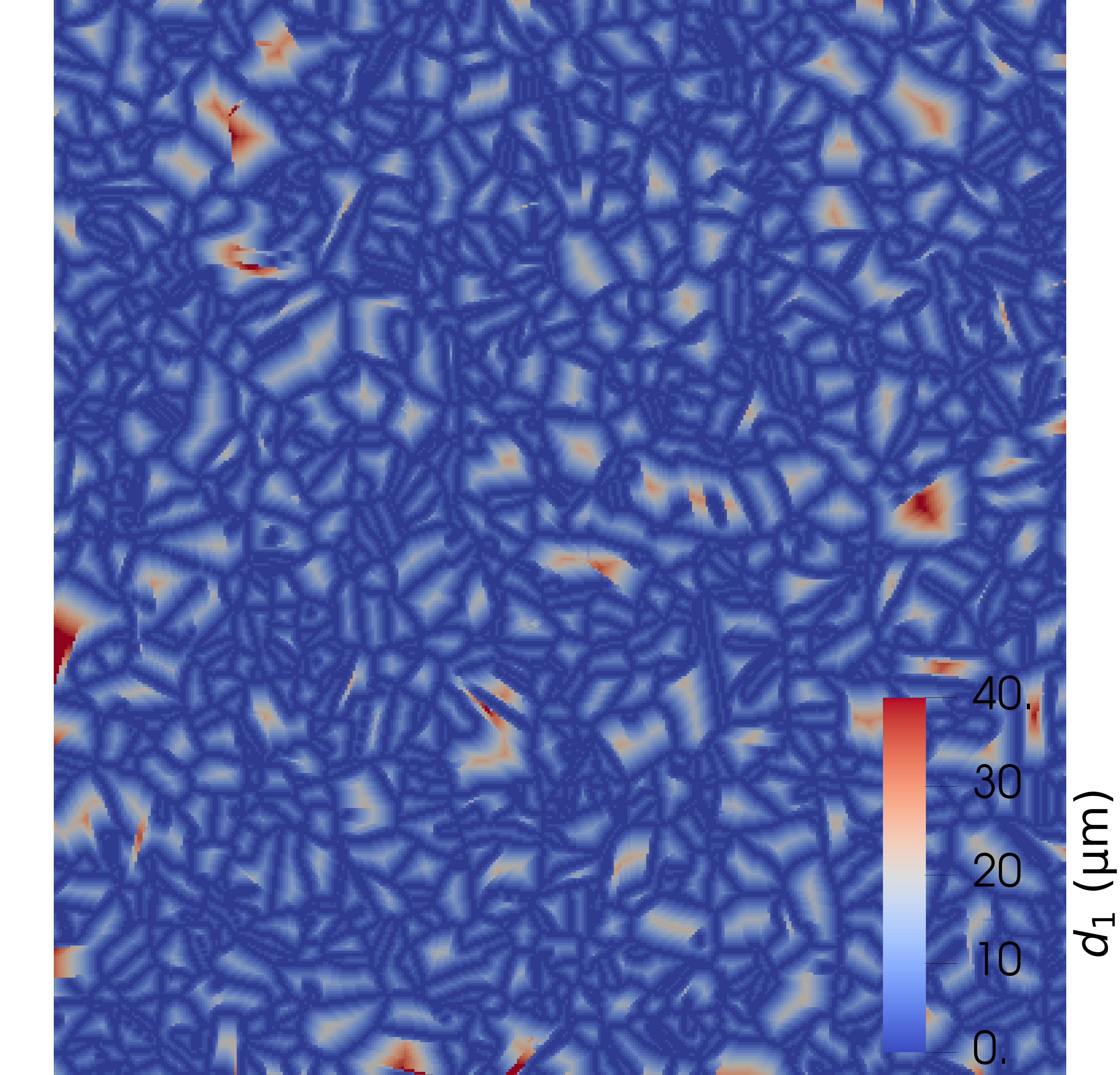}}
\caption{(a) 3D microstructure used for the FFT simulations: colors correspond to the different crystallographic orientations (b) \textcolor{black}{Distance $d_1$ to the closest GB along the slip direction 1.}}
\label{fig8}
\end{figure}

The numerical simulations have been performed using the 3D reconstructed microstructure obtained experimentally (Fig.~\ref{fig8}a), using the Fast Fourier Transform (FFT) solver \texttt{AMITEX\_FFTP} \cite{Amitex}. The FFT method relies on regular grids to compute the mechanical equilibrium based on Lippmann-Schwinger equations \cite{moulinecsuquet}.

\vspace{-0.2cm}

The mechanical behavior of all voxels corresponds to the constitutive equations described in the previous section. The parameters of these equations are summarized in Tab.~\ref{tab2}, and the crystallographic orientation of each voxel is given by the experimental values. Only the irradiated layer is considered, with constant materials parameters, thus neglecting the effect of the dose profile along the thickness. The distance $d_b^{\al}$ in Eq.~\ref{eq_rd2} is shown in Fig.~\ref{fig8}b, using the simplification that the distance to the nearest GB \textcolor{black}{is computed in each plane separately.} For each voxel and slip system $\alpha$, the Luster-Morris parameters $N_{\alpha \beta}$ are computed, where $\beta$ corresponds to the slip system of the adjacent grain, and Eq.~\ref{Ks} is used to define $K_S^{\alpha}$. FFT methods only allow to apply macroscopic (volume-average) strain / stress (or mixed) loading conditions on periodic structures. Uniaxial stress loading conditions are applied:
\begin{equation}
  \frac{1}{V}\int_V \bm{\sigma} dV =  \Sigma_{xx} \bm{e}_x \otimes \bm{e}_x
\end{equation}
where the axis $x$ corresponds to the loading axis during the \textcolor{black}{SSRT}. Due to the fact that the microstructure is not periodic along the in-plane directions, such condition might create some artefacts close to the boundary. In addition, to simulate the free-surface effect, a stress-free layer of one voxel is added at the surface. Macroscopic deformation gradient $F_{xx}$ is applied along the $x$ axis up to 4\% strain, as in the \textcolor{black}{SSRT}. All simulations have been performed using the full 3D microstructure ($\mathrm{800 \mu m}\ \mathrm{x}\ \mathrm{800 \mu m}\ \mathrm{x}\ \mathrm{20 \mu m}$) with an in-plane discretization of $2 \mu$m and through thickness discretization of $0.5 \mu$m, corresponding to 6M voxels. \textcolor{black}{Additional simulations presented in Appendix~C have shown that coarser discretizations or considering a larger thickness does not affect the conclusions presented hereafter.} Simulations have been performed using $K_s^0 \in [0;5]$ to evaluate the effect of slip transmission modelling on GB local stresses. \textcolor{black}{The first case corresponds to fully transparent GB, the later to the value used in \cite{HAOUALA2020102600}.}

\subsection{Numerical results}

\begin{figure}[H]
  \centering
\subfigure[]{\includegraphics[height = 6.5cm]{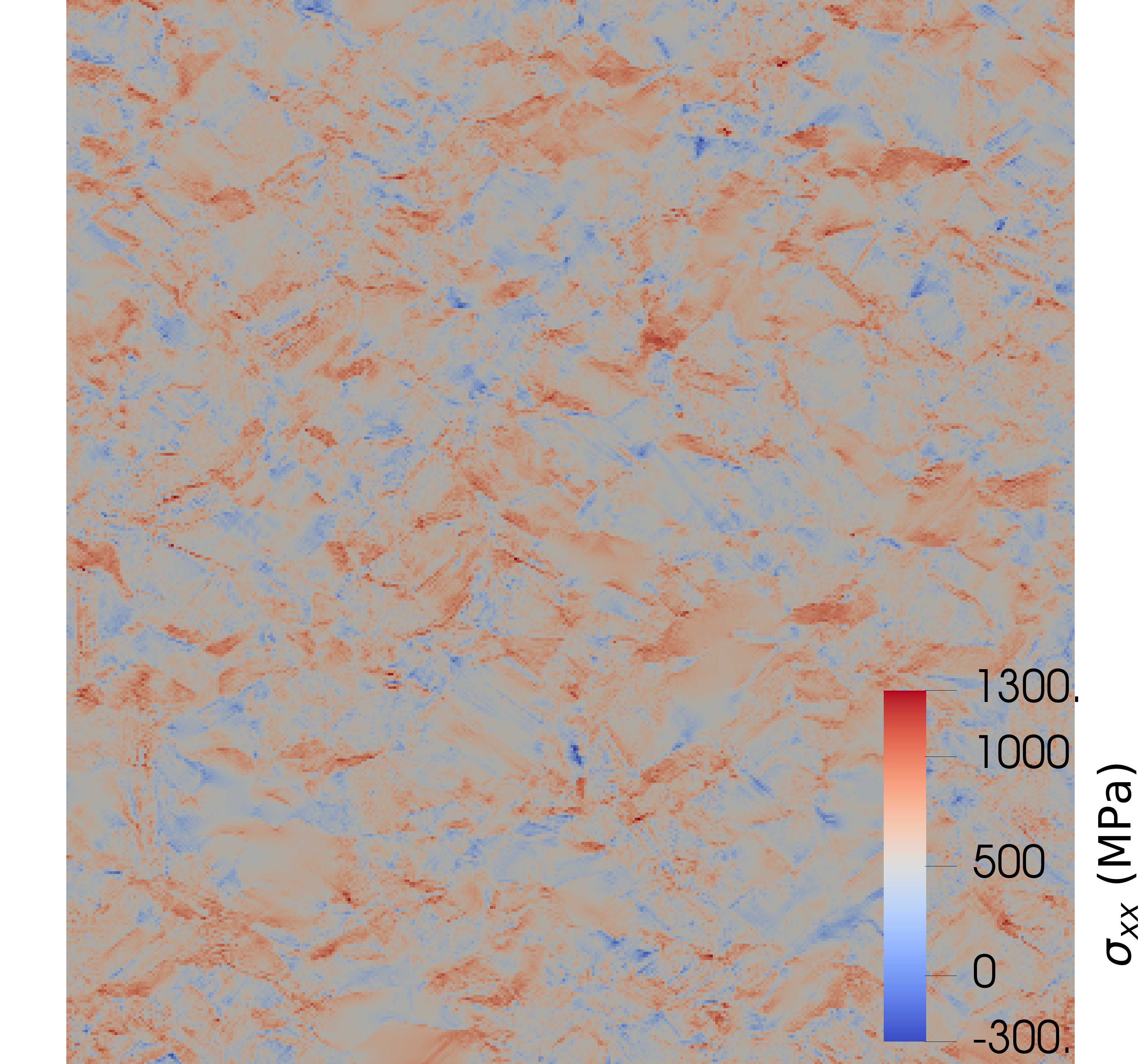}}
\hspace{-0.cm}
\subfigure[]{\includegraphics[height = 6.5cm]{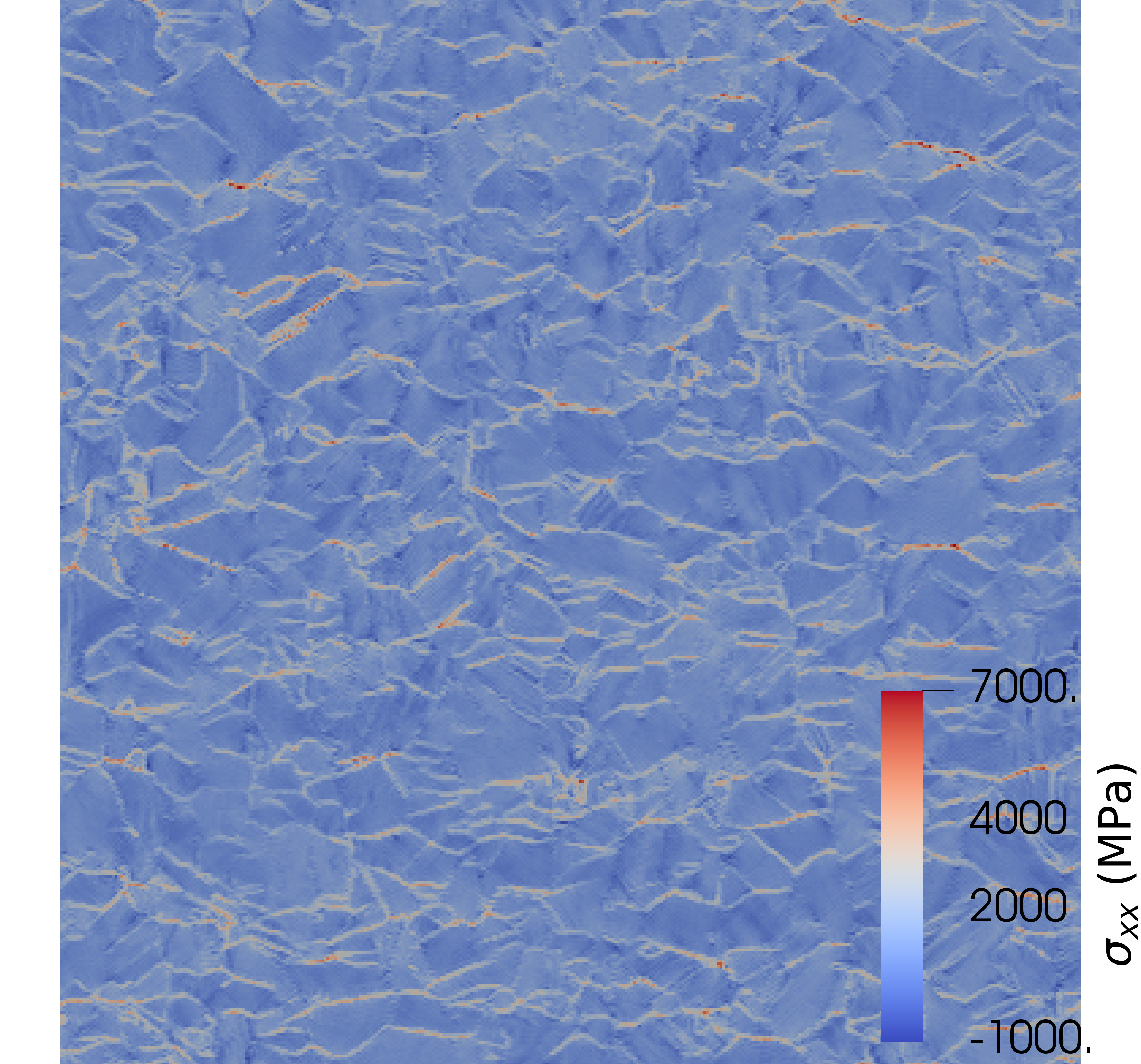}}
\caption{Stress $\sigma_{xx}$ field at the free surface of the aggregate for an applied strain of $4\%$, for (a) $K_s^0 = 0$ and (b) \textcolor{black}{$K_s^0 = 5$}}
\label{fign1}
\end{figure}

\begin{figure}[H]
  \centering
\subfigure[]{\includegraphics[height = 6.5cm]{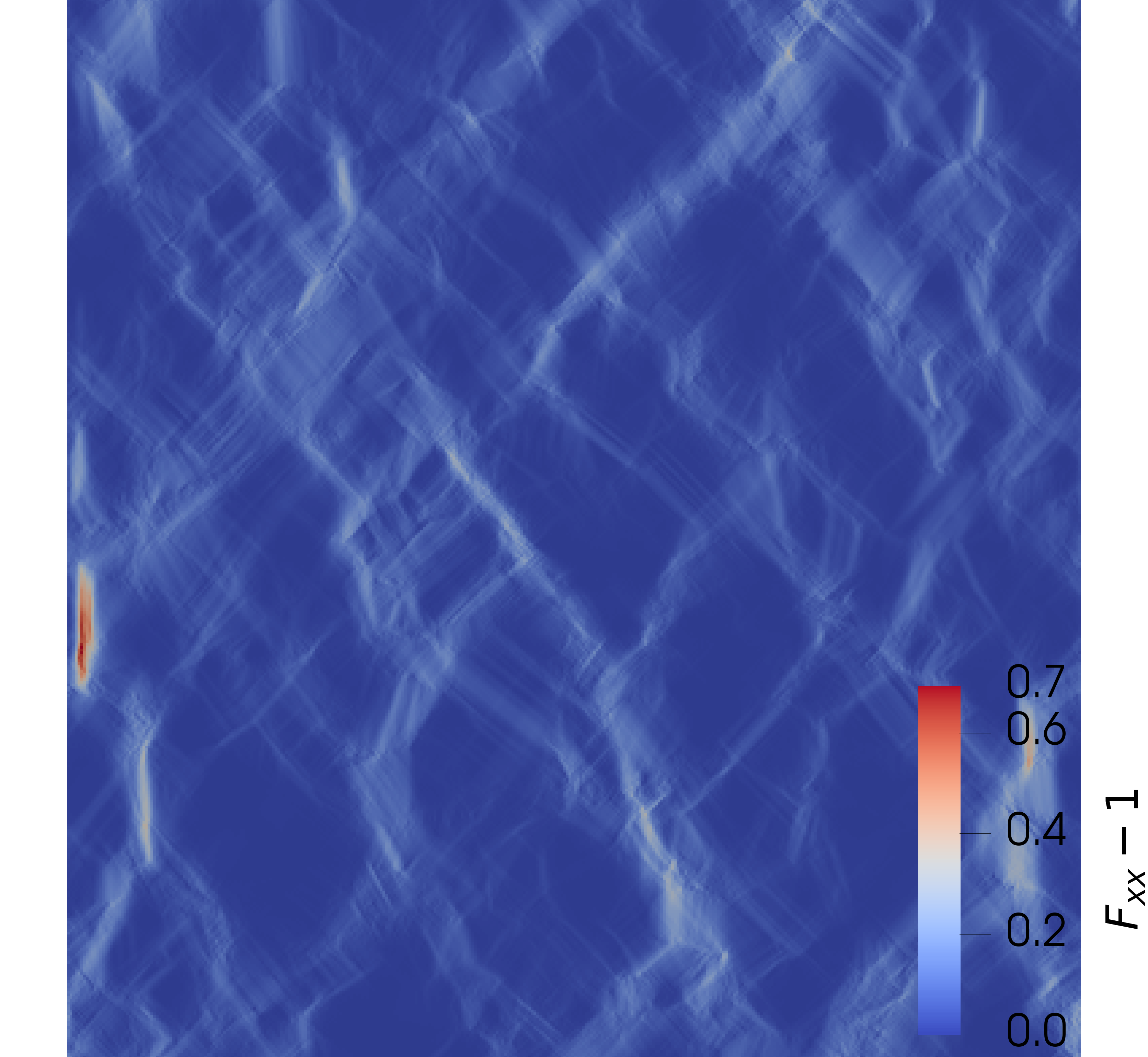}}
\hspace{-0.cm}
\subfigure[]{\includegraphics[height = 6.5cm]{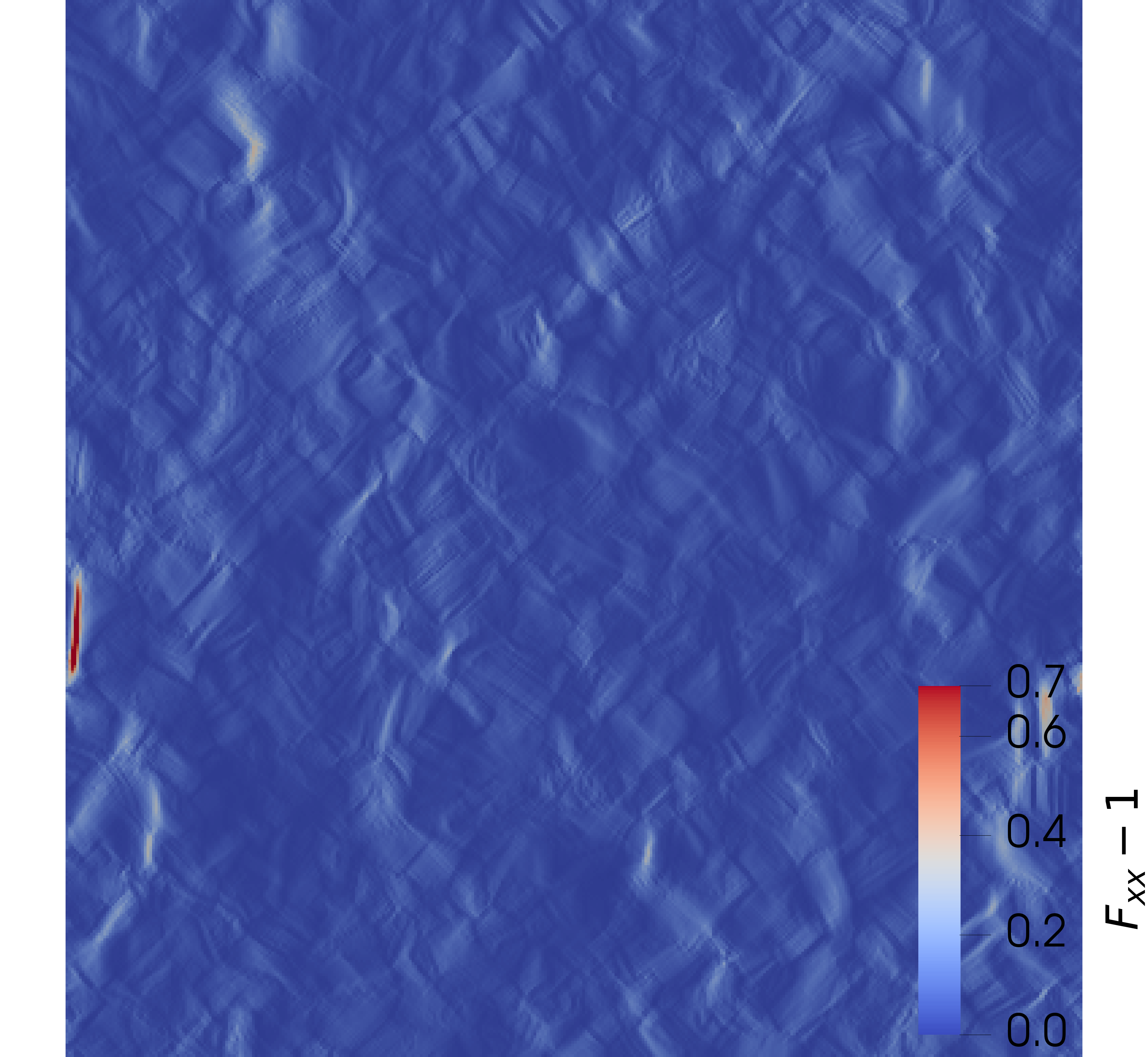}}
\caption{Deformation gradient $F_{xx}$ field at the free surface of the aggregate for an applied strain of $4\%$, for (a) $K_s^0 = 0$ and (b) \textcolor{black}{$K_s^0 = 5$}}
\label{fign2}
\end{figure}

Simulations have been performed on the \texttt{Cobalt} supercomputer (CCRT/CEA) corresponding to a typical runtime of about 48h on 1624 processors for each simulation. Typical results at the free surface of the aggregate are shown on Figs.~\ref{fign1} and \ref{fign2} for stress and strain fields, respectively. For $K_s^0=0$, \textit{i.e.}, fully transparent GB, grain to grain variations of local stress (Fig.~\ref{fign1}a) can be observed due to the different crystallographic orientations. Grain boundaries do not appear to have higher stresses than the interior of adjacent grains, consistent with the absence of any GB modelling. Interesting features appear on strain field as shown on Fig.~\ref{fign2}a. At the scale of the aggregate, localization bands spanning multiple grains emerge, while at the grain scale, additional localization bands are present.

\begin{figure}[H]
\centering
\subfigure[]{\includegraphics[height = 5.cm]{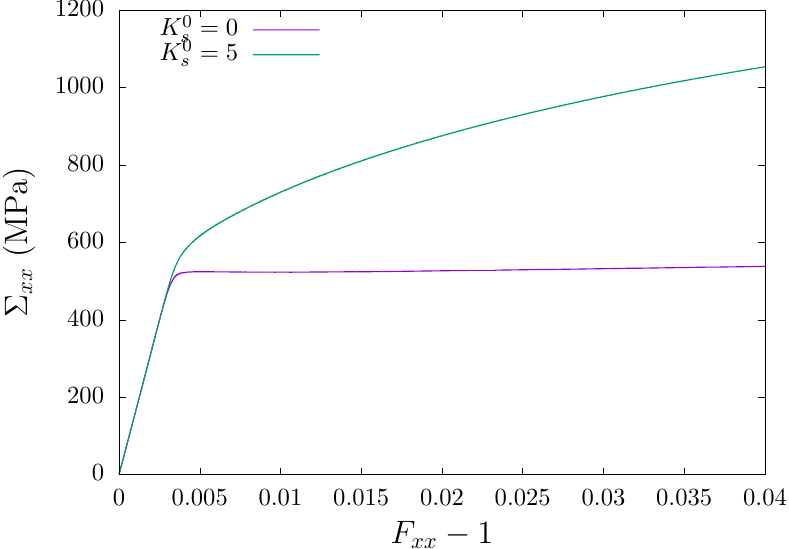}\vspace{0.5cm}}
\hspace{-0.cm}
\subfigure[]{\includegraphics[height = 5.cm]{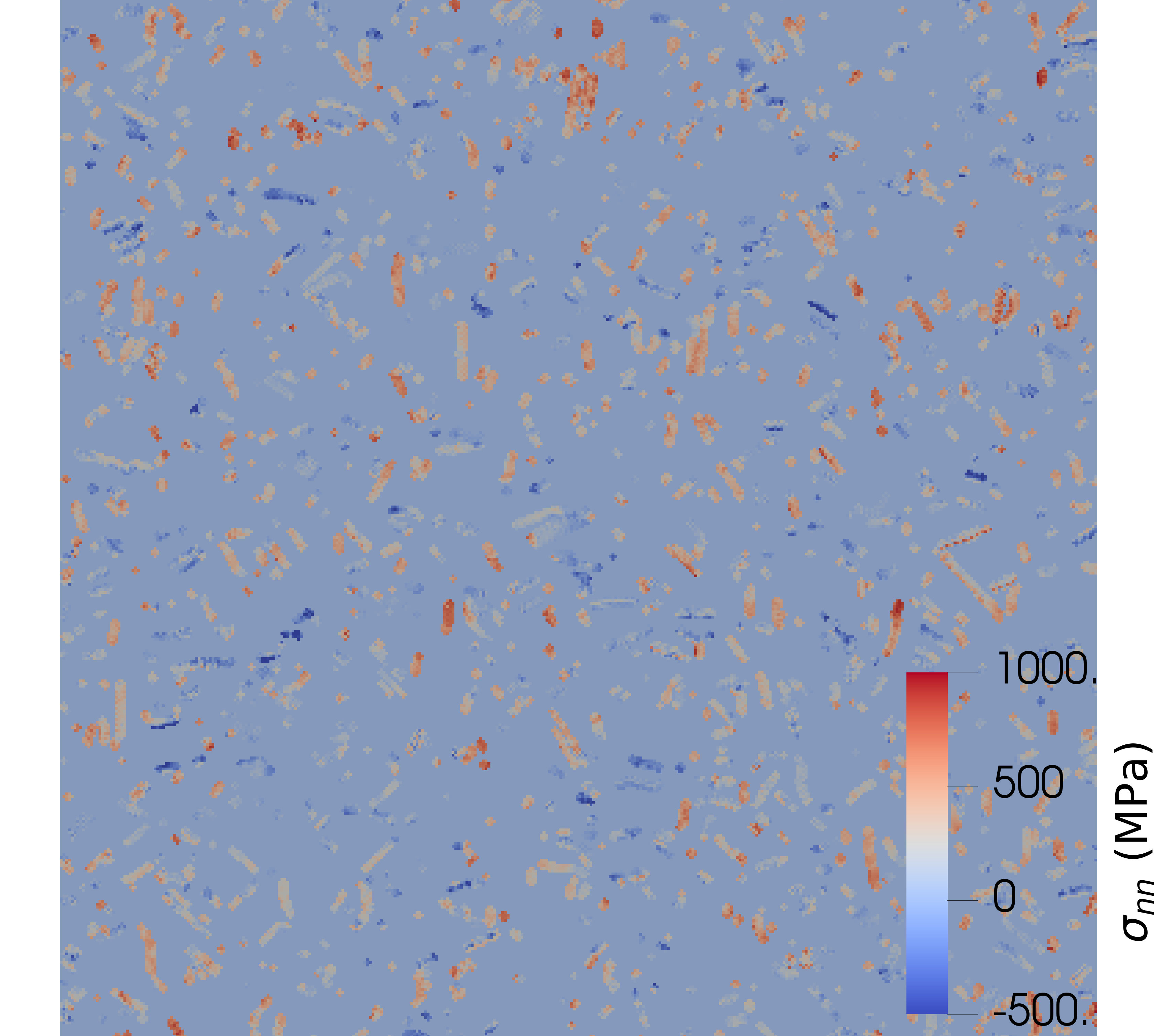}}
\caption{(a) Aggregate stress-strain curves (b) Intergranular normal stresses $\sigma_{nn}$ at the free surface of the 3D microstructure for $K_s^0=0$}
\label{fig9}
\end{figure}

The root cause of these localizations is crystal scale (through the removal of irradiation defects) and aggregate scale (as shown on Fig.~\ref{fig9}a and discussed hereafter) softening mechanical behavior. Intragranular localization bands are expected to be mesh dependent, as no regularization is used in the model. Results are drastically different using GB modelling, as shown on Figs.~\ref{fign1}b and~\ref{fign2}b for \textcolor{black}{$K_s^0 = 5$}. Stresses are higher at GB than in the interior of the grains due to the absence of slip transmission and thus multiplication of dislocations close to these GB, as modelled by Eq.~\ref{eq_rd2}. Concurrently, strains are higher in grains' interior than close to the boundary, as shown on Fig.~\ref{fign2}b. Still some intragranular localization bands can be observed, but no localization at the aggregate scale.

The aggregate stress-strain curves are shown on Fig.~\ref{fig9}a. For $K_s^0 = 0$, an almost perfectly plastic behavior is obtained, which is consistent with a 2dpa 304L mechanical behavior \cite{HURE2016231}. A slight softening is observed just after yielding, followed by a hardening behavior with a small hardening modulus. The overall stress magnitude is lower than the one reported in \cite{HURE2016231} using the same constitutive equations on a 3D aggregate due to the small thickness of the aggregate used in this study and hence where plane stress conditions prevail. As discussed before, it has been verified that decreasing the thickness does indeed change the macroscopic stress-strain curves, but not the values of the stresses at the free surface \textcolor{black}{(Appendix~C)}, which is the main interest for cracking initiation. For $K_s^0 > 0$, a strong hardening behavior is observed due to the accumulation of dislocations at grains boundaries where slip transmission is unlikely \cite{HAOUALA2020102600}. As expected, the stress-strain curves for \textcolor{black}{$K_s^0 = 5$} are not consistent with a 2dpa austenitic stainless steels, as the other hardening parameters of the constitutive equations have been calibrated assuming $K_s^0 = 0$. In order to get the desired almost perfectly plastic macroscopic behavior, a stronger softening due to the removal of irradiation defects should be considered, that will lead to more heterogeneous intragranular deformation behavior. Such calibration will be considered in future studies.

Local stresses at the free surface  along with GB locations and normals are used to compute intergranular normal stresses for a macroscopic applied strain value of $4\%$ that corresponds to the maximal value at the end of the \textcolor{black}{SSRT}\footnote{Analysis of the numerical data at other timesteps does not change the results reported hereafter.} in PWR environment:
\begin{equation}
  \sigma_{nn} = \bm{n} \cdot {\bm{\sigma}} \cdot \bm{n}
\end{equation}
A typical result of the computation of intergranular normal stresses is shown on Fig.~\ref{fig9}b, where the magnitude of $\sigma_{nn}$ can be seen to strongly depend on GB positions. These results are used to compute cumulative distribution functions of $\sigma_{nn}$ for both uncracked and cracked GB as well as to evaluate the effect of the GB modelling through the parameter $K_s^0$. \textcolor{black}{As $\Sigma_3$ GB have been shown in the literature and in this study to be resistant to cracking, distributions are computed only on non $\Sigma_3$ GB.}

\begin{figure}[H]
\centering
\subfigure[]{\includegraphics[height = 4.5cm]{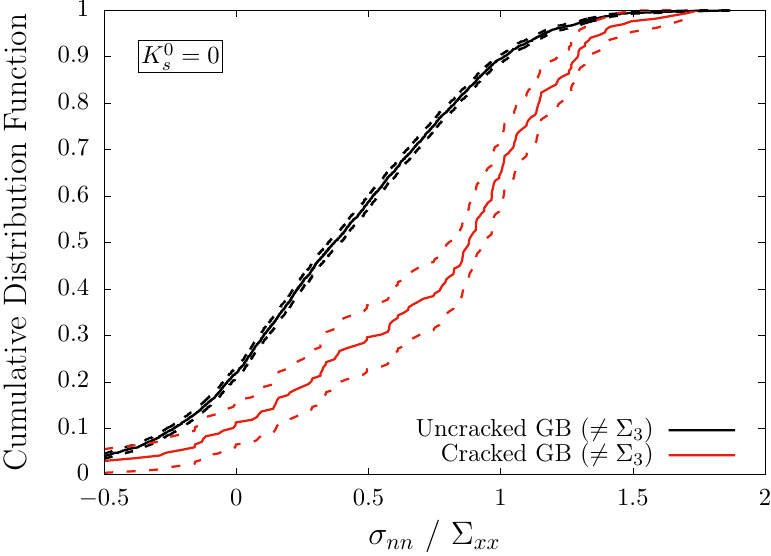}}
\hspace{0.5cm}
\subfigure[]{\includegraphics[height = 4.5cm]{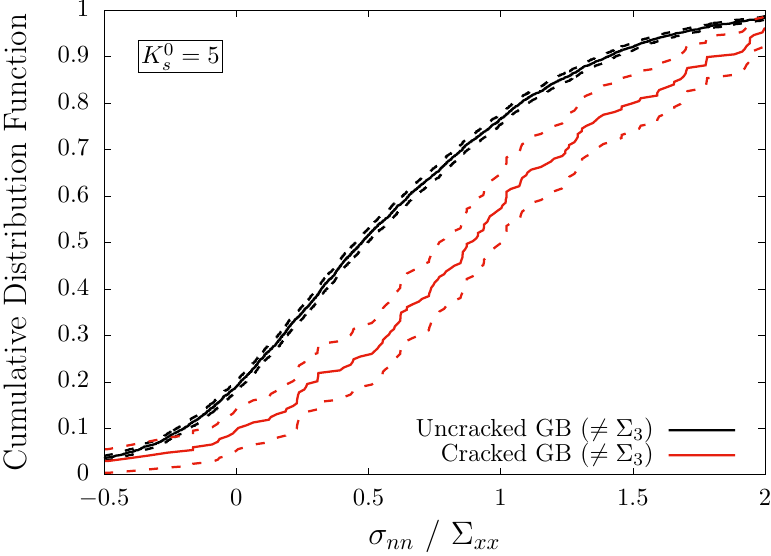}}
\caption{\textcolor{black}{Cumulative distribution functions of normalized intergranular normal stress  $\sigma_{nn} / \Sigma_{xx}$ for uncracked (in black) and cracked (in red), for (a) $K_s^0 = 0$ and (b) $K_s^0 = 5$}}
\label{fign3}
\end{figure}

For $K_{s}^0 = 0$, \textit{i.e.}, fully transparent GB, normalized intergranular normal stress at cracked GB are found to be \textcolor{black}{higher} than for uncracked GB (Fig.~\ref{fign3}a). This observation is consistent with the analysis of the experimental data shown in Fig.~\ref{fig55} where larger values of the normal component along the loading axis was found, which has been related to higher intergranular normal stresses. \textcolor{black}{The difference is rather significant: about 60\% of cracked GB have for example intergranular normal stresses $\sigma_{nn} / \Sigma_{xx} \geq 0.8$, compared to about 20\% of uncracked GB.}
\textcolor{black}{For $K_{s}^0 = 5$ (Fig.~\ref{fign3}b)}, a shift towards the higher $\sigma_{nn}$ values is observed for both uncracked and cracked GB, \textcolor{black}{showing the effectiveness of such GB modelling}.
\textcolor{black}{Differences are still observed between uncracked and cracked GB.} However, for the current model, it is observed to affect similarly uncracked and cracked GB, showing the difference pointed out in Fig.~\ref{figsigma3}b is not sufficient to increase local stresses at cracked GB. \textcolor{black}{The results shown on Fig.~\ref{fign3} indicate that cracked GB have statistically higher intergranular normal stresses, but the differences observed are not sufficient to define a cracking criterion. Accounting for GB modelling through Luster-Morris parameter, although found to correlate experimentally with cracking, does not allow to exacerbate the differences.}

\begin{figure}[H]
\centering
\subfigure[]{\includegraphics[height = 4.5cm]{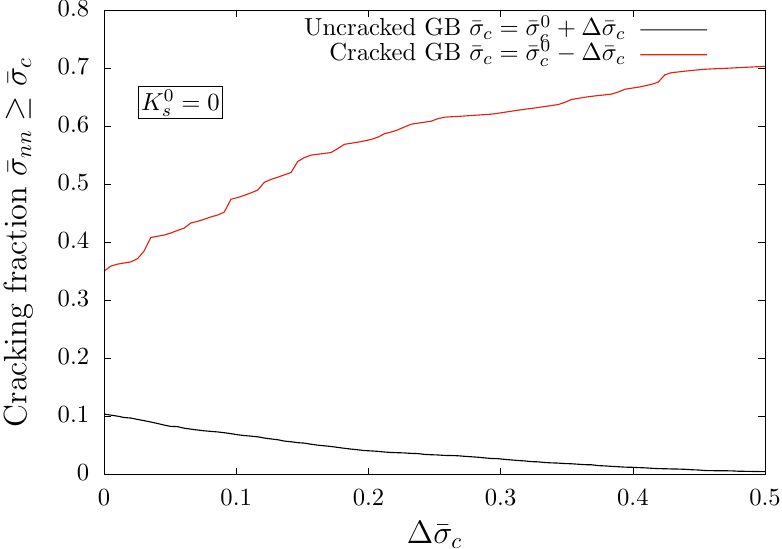}}
\hspace{0.5cm}
\subfigure[]{\includegraphics[height = 4.5cm]{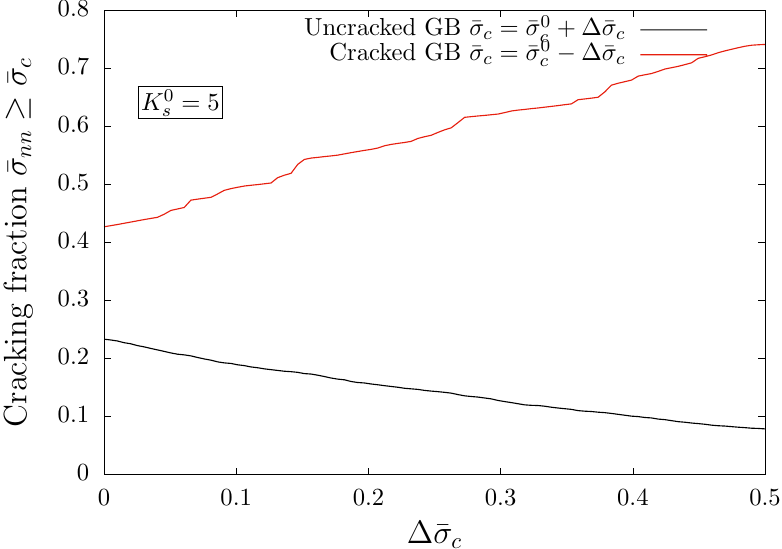}}
\caption{\textcolor{black}{Cracking fraction for uncracked and cracked GB as a function of differential GB strength $\Delta \bar{\sigma}_c$, for $\sigma_c^0 = \Sigma_{xx}$, for (a) $K_s^0 = 0$ and (b) $K_s^0 = 5$}}
\label{fignew5}
\end{figure}

This was somehow expected as the underlying assumption of this analysis is that all (\textcolor{black}{non $\Sigma_3$}) GB have the same strength, which is clearly a strong assumption. It is therefore interesting to assess the effect of considering different GB strengths on cracking predictions, based on these numerical results. \textcolor{black}{Fig.~\ref{fignew5}} shows the percentage of cracking for GB defined as cracked and uncracked based on experimental observations, considering a differential GB strength $\Delta \bar{\sigma}_c$ between uncracked GB and cracked GB. A fully predictive model corresponds to a cracking fraction of 1 for cracked GB and 0 for uncracked GB. Considering a similar strength for all GB ($\Delta \bar{\sigma}_c = 0$ and $\sigma_c$ sets arbitrarily to the macroscopic stress), the difference between cracking fraction is small. Increasing $\Delta \bar{\sigma}_c$ leads to drastic evolutions for the cracking fractions, showing that such micromechanical approach coupled with some models able to predict accurately GB strength may be a promising approach. 

\section{Discussion}
\label{sec4}

The key experimental result of this study is the correlation observed between cracking initiation and the Luster-Morris slip transmission criterion. This result allows to go beyond correlations based on slip discontinuity observed after the tests, and opens the way for a micromechanical approach for IASCC that may not need to account explicitly for dislocation channelling phenomenon. The strengths and weaknesses of the current micromechanical approach are now detailed.

\subsection{Towards a micromechanical approach for IGSCC}

Crystal plasticity simulations on realistic polycrystalline aggregates allow evaluating local stresses / strains fields accounting for the geometry - due grain shapes - and anisotropy - due to the crystallographic orientations in a way which is both physical and efficient, using the ability of massively parallel FFT-based solvers to deal with large-scale simulations. This allows to account naturally for both the effects of GB normals and grain-grain interactions on intergranular stresses, which is clearly the main strength of such approach as demonstrated by the previous results. The main weakness lies in the crystal scale constitutive equations and GB modelling. First, even if the equations used to describe hardening and evolutions of dislocations and defects are physically based, the calibration and validation of the parameters remains incomplete due to the lack of experimental results on single crystals. This is particularly true for irradiated austenitic stainless steels where only few data are available in the literature \cite{JIN2016155,VO2017336,PACCOU201956}, most of them showing size effects due to the small-scale samples used. While some parameters of the models described in \cite{HURE2016231} are taken from lower-scale simulations (\textit{e.g.}, Discrete Dislocation Dynamics for the coefficients of the interaction matrix), others (especially those related to hardening) are calibrated against polycrystalline aggregates results, which does not ensure that the constitutive equations would lead to satisfactory results for single crystals. Secondly, although the modelling of GB effect proposed in \cite{HAOUALA2020102600} and used in this study is both physically-based and numerically efficient, it remains to be supported by dedicated experimental observations of slip transmission in irradiated stainless steels, as well as by experimental local stresses measurements. Although requiring a lot of experimental observations, these two weaknesses could be overcome in the near future. The modelling of intragranular strain localization in polycrystalline aggregates is a bigger challenge. It has been shown how the impingement of a well-defined slip band on a given GB could lead to fracture \cite{sauzay}, but applying the same criterion requires to simulate all slip bands appearing at the crystal scale. Gradient enhanced crystal plasticity constitutive equations are needed to regularize strain localization and to ensure mesh independence in numerical simulations. Moreover, as the softening behavior is a necessary condition for strain localization, improving the calibration of constitutive equations is also necessary. Besides, although numerical tools are available in the literature to perform large-scale simulations exhibiting regularized intragranular strain localization \cite{SCHERER2019103768,MARANO2019262}, as observed experimentally for irradiated austenitic stainless steels, how to control the distance between slip bands remains unknown (although theoretical models are developed to predict such distance \cite{Gururaj2015}). In the meantime, accounting for slip transmission through the model proposed in \cite{HAOUALA2020102600}, \textcolor{black}{once properly calibrated to affect only GB prone to cracking}, is a promising mesoscopic approach to detect GB where slip continuity is unlikely and thus more prone to cracking, without having to explicitly account for dislocation channelling.

All the aforementioned strengths and weaknesses of the micromechanical approach deal with accurate predictions of local stresses. Intergranular cracking will of course depends also on GB strength that may depend on irradiation, oxidation time, GB type. The analysis of the experimental and numerical results presented in the previous sections make it clear that a high local intergranular stress is not a sufficient condition for cracking, and that differential GB strength may have an important effect on cracking prediction (Fig.~\ref{fignew5}). GB strength and intergranular oxidation are discussed in the next section.

\subsection{Grain boundary oxidation}

\textcolor{black}{The strength of a grain boundary, defined in the previous section as the critical normal stress above which a GB will fracture, is not an intrinsic property. It should be considered as an effective property that depends on GB type - as shown for example by the fact that $\Sigma_3$ GB are considered to be resistant to cracking \cite{GERTSMAN20011589} - and on the environmental embrittlement. As such,} grain boundary oxidation is expected to play a key role in IGSCC. For some Nickel-based alloys, dedicated experiments allowed to quantify intergranular oxidation in high temperature aqueous environment \cite{duhamel} \textcolor{black}{as well as the dramatic degradation of GB strength with oxidation \cite{fujii}.}  However, even for unirradiated austenitic stainless steels, experimental quantification of intergranular oxidation in PWR environment is scarce. In \cite{MATTHEWS2017175}, TEM observations performed on several GB showed that oxidation strongly depends on the GB considered, which could explain the fact that for two GB having the same estimated local stresses, only one of them (or none) will fail due to different oxidation states. For irradiated austenitic stainless steels, quantification of intergranular oxidation in PWR environment is an identified research gap. \textcolor{black}{Indirect observations are available based on the effect of GB type on irradiation-induced segregation. In particular, $\Sigma_3$ GB have been shown to be resistant to segregation \cite{DUH2000198}.} Dedicated experimental programs are definitely required to quantify, as a function of oxidation time, irradiation level and GB type, intergranular oxidation and correlations between cracking and microchemistry. Although experimental techniques - such as TEM on FIB samples - are available, it will require a huge amount of time, but appears unavoidable to improve the micromechanical approach of IGSCC. Regarding the effect of GB type on intergranular oxidation in austenitic stainless steel, it has been proposed recently a model based on the definition of an Atomic Packing Density (APD) that showed a promising correlation with intergranular corrosion in highly corroding environment \cite{AN2018297}. For the experimental data obtained in this study, APD was computed for each grains, and the minimal value of the two grains adjacent to a GB was used as the GB APD. However, no difference is observed between uncracked and cracked GB, indicating that GB APD, as described in \cite{AN2018297}, may not be able to predict GB oxidation / strength of austenitic stainless steels in PWR environment.

\section{Conclusion and Perspectives}

A micromechanical analysis of IGSCC of an irradiated austenitic stainless steel has been proposed, based on the reconstruction (through sequential polishing and 2D EBSD) of the 3D microstructure of a 304L proton-irradiated sample tested in PWR environment. This analysis allows to go beyond the analysis often reported in the literature based only on 2D observations, in particular with respect to the effect of GB normal and slip transmission criteria. Moreover, the large volume considered (and thus large number of cracks) has made possible a statistical analysis of the data. A correlation is observed between intergranular cracking and GB well-oriented with respect to the mechanical loading applied during the IGSCC test. In addition, evaluation of several slip transmission criteria has shown a correlation between the Luster-Morris parameter - involving only the crystallographic orientations on both sides of the GB - and cracking. Interestingly, this slip transmission criterion has been put forward for other FCC materials in recent studies showing its ability to predict slip (dis-)continuity.
Micromechanical simulations based on the reconstructed 3D microstructure and crystal plasticity constitutive equations modified to account for slip transmission at GB have been performed. The two main outcomes of these simulations are first higher local stresses for cracked GB, and secondly evidence that considering differential GB strength has a strong effect on cracking predictions. In addition, GB modelling described in \cite{HAOUALA2020102600} appears numerically efficient and effective to increase local stresses where slip transmission is unlikely, and a promising mesoscopic way of accounting for the observed dependence of cracking on slip discontinuity without having to describe explicitly dislocation channels. These simulations allow to point out the main research gap that lies in finding the dependence of GB strength on GB type and oxidation,  setting the path for future studies.\\

\noindent
\textbf{Acknowledgements}\\

The authors gratefully acknowledge financial support provided by Slovenian Research
Agency (grant P2-0026) and French Atomic Energy Commission.\\

\noindent
\textbf{CRediT Author statement}\\

\noindent
\textbf{Disen Liang}: Investigation, Software, Formal analysis, Writing - Review \& Editing \textbf{J\'er\'emy Hure}: Investigation, Software, Formal analysis, Writing - Original draft, Supervision, Conceptualization \textbf{Arnaud Courcelle}: Investigation, Writing - Review \& Editing \textbf{Samir El Shawish}: Investigation, Software, Writing - Review \& Editing \textbf{Beno\^it Tanguy}: Conceptualization, Writing - Review \& Editing 

\section*{Appendix A: 3D EBSD reconstruction}

\begin{figure}[H]
\centering
\includegraphics[height = 5.0cm]{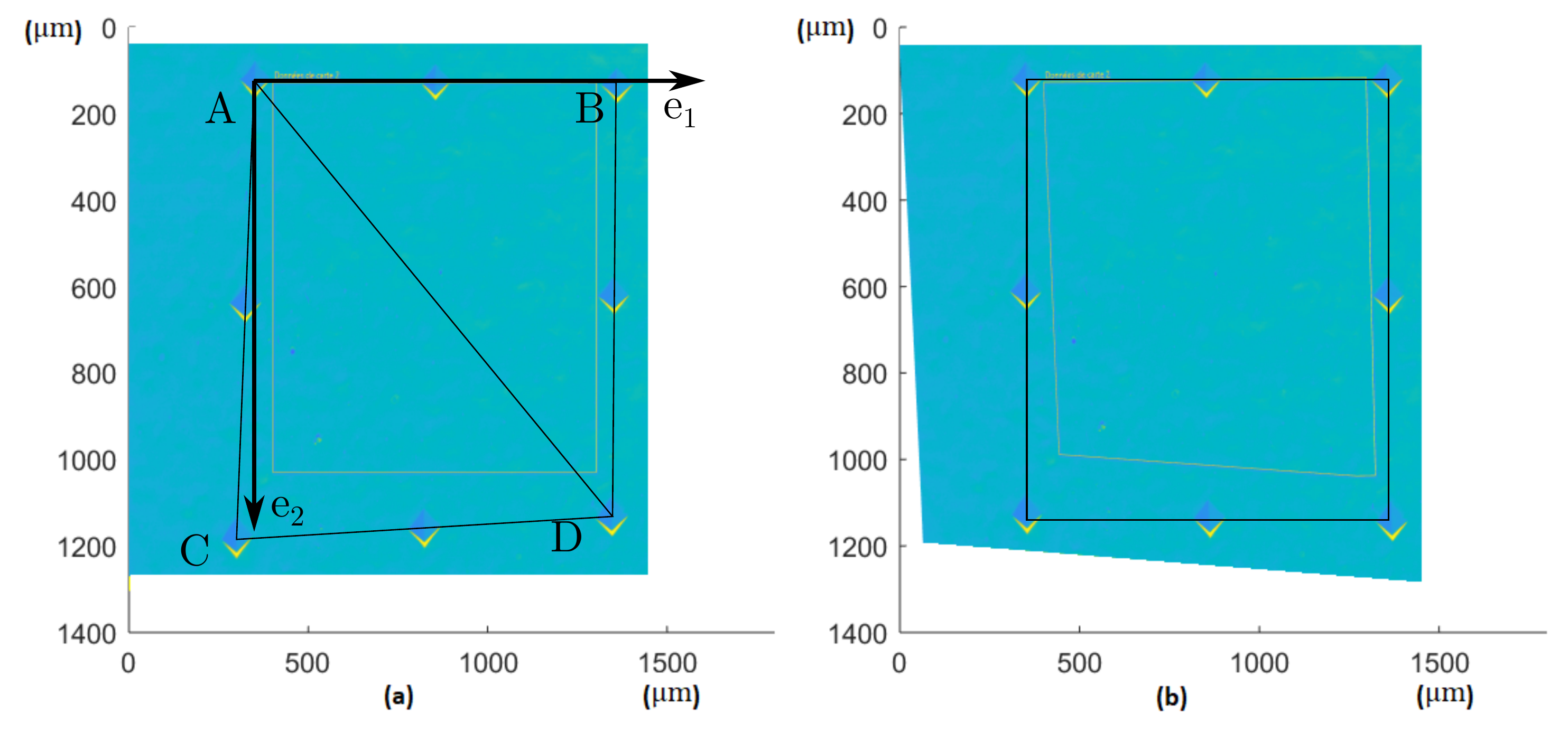}
\caption{(a) SEM image of the area identified by the indents in the EBSD configuration (70$^{\circ}$ tilt angle) (b) Correction of the SEM image}
\label{figAA}
\end{figure}

This section provides the technical details regarding the EBSD data corrections and 3D EBSD reconstruction. For each EBSD map, the crystallographic phases and orientations are obtained in the frame $\{ e_1,e_2\}$ shown in Fig.~\ref{figAA}a. The Euler angles $\Psi_i$ are defined at locations:
\begin{equation}
  \Psi_i\left(x,y \right) = \Psi_i(\textbf{X})
\end{equation}

\begin{figure}[H]
\centering
\includegraphics[height = 2.5cm]{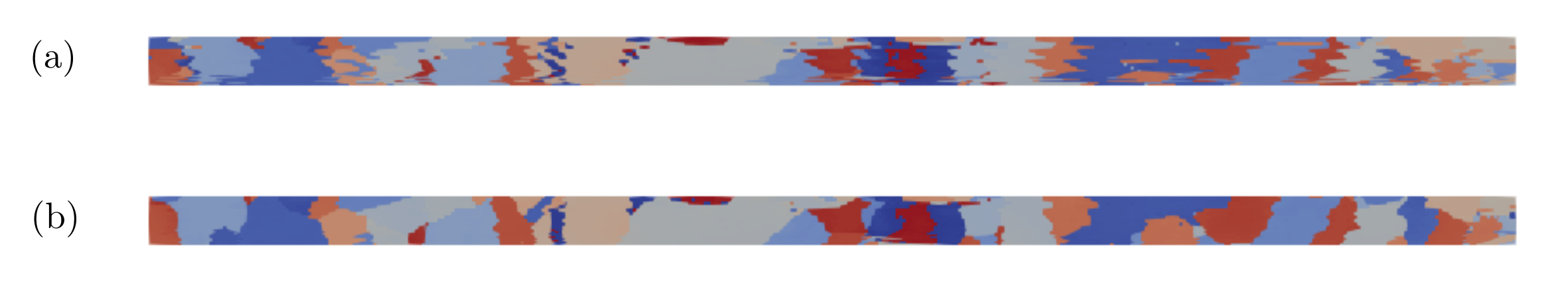}
\caption{Stacking of successive 2D EBSD maps to reconstruct the 3D microstructure, (a) without and (b) with the use of Eq.~\ref{argmin}}
\label{figAAA}
\end{figure}

Corrections of the coordinates are necessary to remove image distortions such as the positions of the indents:
\begin{equation}
  \textbf{X}^A = \left(
    \begin{array}{c}
      0 \\
      0
    \end{array}
    \right)
    \ \ \ \ \
  \textbf{X}^B = \left(
    \begin{array}{c}
      \alpha_1 \\
      0
    \end{array}
    \right)
    \ \ \ \ \
      \textbf{X}^C = \left(
    \begin{array}{c}
      \alpha_2 \\
      \alpha_3
    \end{array}
    \right)
    \ \ \ \ \
      \textbf{X}^D = \left(
    \begin{array}{c}
      \alpha_4 \\
      \alpha_5
    \end{array}
    \right)
    \ \ \ \ \
\end{equation}
go back to their initial positions and forming a square of size 1 (Fig.~\ref{figAA}b):
\begin{equation}
  \textbf{X}^A_0 = \left(
    \begin{array}{c}
      0 \\
      0
    \end{array}
    \right)
    \ \ \ \ \
  \textbf{X}^B_0 = \left(
    \begin{array}{c}
      1 \\
      0
    \end{array}
    \right)
    \ \ \ \ \
      \textbf{X}^C_0 = \left(
    \begin{array}{c}
      0 \\
      1
    \end{array}
    \right)
    \ \ \ \ \
      \textbf{X}^D_0 = \left(
    \begin{array}{c}
      1 \\
      1
    \end{array}
    \right)
    \ \ \ \ \
\end{equation}
which allows to define piecewise linear transformations in the triangles (ACD) and (ABD) and to determine the real position of the Euler angles measurements:
\begin{equation}
  \textbf{X}_0 = \textbf{{M}} \cdot \textbf{X} \ \ \ \ \ \ \ \ \ \ \Psi_i(\textbf{X}_0) = \Psi_i \left(\textbf{{M}} \cdot \textbf{X} \right)
\end{equation}
The coefficients $\alpha_i$ corresponding to the positions of the Vickers indents are measured on SEM images. However, as shown on Fig.~\ref{figAAA}a, applying separately corrections to each 2D EBSD map does not prevent for 3D reconstruction artefacts. Corrections of the positions of each Vickers indents $\Delta \alpha_i$ are thus allowed, and determined through maximizing the cross-correlation of the first Euler angles $\phi_1$ between the current EBSD map and the previous one:
\begin{equation}
  \Delta \alpha_i^n = \mathrm{argmax}\left( \phi_1^n \star \phi_1^{n-1}    \right)
  \label{argmin}
\end{equation}
where $\star$ denotes the convolution product. Derivative-free Nelder-Mead algorithm is used to compute Eq.~\ref{argmin}, allowing to successfully align successive EBSD maps, as shown in Fig.~\ref{figAAA}b. All computations have been performed using \texttt{Matlab} software.

\begin{figure}[H]
\centering
\subfigure[]{\includegraphics[height = 4.5cm]{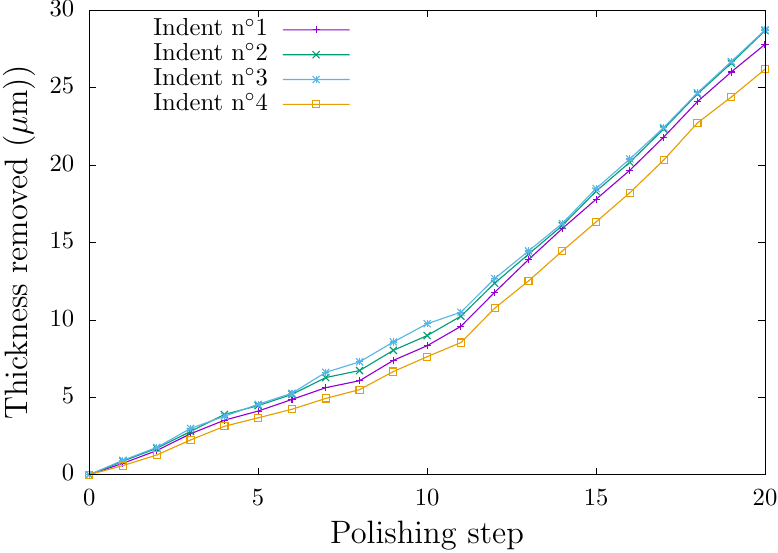}}
\hspace{0.5cm}
\subfigure[]{\includegraphics[height = 4.5cm]{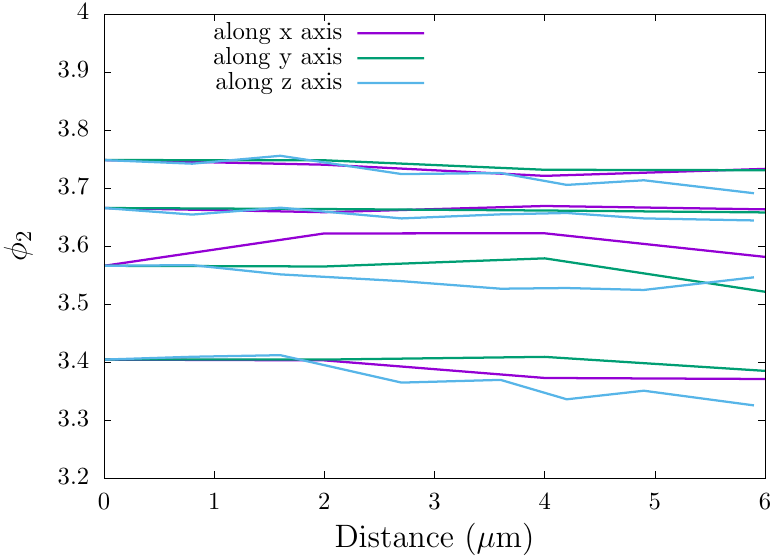}}
\caption{\textcolor{black}{(a) Evolutions of the thickness removed measured at the four Vickers indents locations as a function of the polishing step (b) Evolutions of the Euler angle $\phi_2$ along the three axis, starting from points located at the surface of the reconstructed microstructure in the middle of large grains}}
\label{figAAA0}
\end{figure}

\textcolor{black}{The reconstruction algorithm allows to correct in-plane misalignments of successive EBSD scans, but no correction is done regarding potential out-of-plane misalignments. Fig.~\ref{figAAA0}a shows the evolutions of the thickness removed measured at the four Vickers indents located at the corners of a $l = 1$mm square. The maximum absolute difference is about $\Delta h = 2\mu$m, which corresponds to an angle of $\theta \approx \Delta h / l \approx 0.1^{\circ}$ with respect to the initial plane. Thus, out-of-plane misalignment induced by the polishing procedure is negligible.}

\textcolor{black}{Still, misalignment coming from the incorrect mounting of the sample in the SEM sample holder needs to be quantified. Points are selected at the surface of the reconstructed microstructure (Fig.~\ref{figAAA}b) in the middle of large grains, and the evolutions of the Euler angles along the three axis are shown on Fig.~\ref{figAAA0}b. The variations along the thickness (z-axis) are of the same order as the in-plane variations, indicating that the out-of-plane misalignment is negligible.}

\section*{Appendix B: Detection of grain boundaries}

This section provides a more detailed technical description of the
 approach employed in this study to reconstruct free-surface GB
of a polycrystalline sample. A focus is set to the calculation of
in-plane and out-of-plane GB slopes given the two EBSD images obtained
on two parallel surfaces initially stacked atop each other in a
specimen.The first step consists of identifying all GB voxels on a top plane
grid. A voxel is defined to be a GB voxel if there is exactly one
additional color recognized (within some prescribed tolerance) in its
immediate neighborhood composed of four voxels (up, down, left, right
neighbors). In this way, each identified GB voxel is characterized by
the corresponding color pair. However, if two or more additional
colors are found in the neighborhood, a voxel is defined to be a
triple point voxel.

\begin{figure}[H]
  \centering
\subfigure[]{\includegraphics[width=4cm]{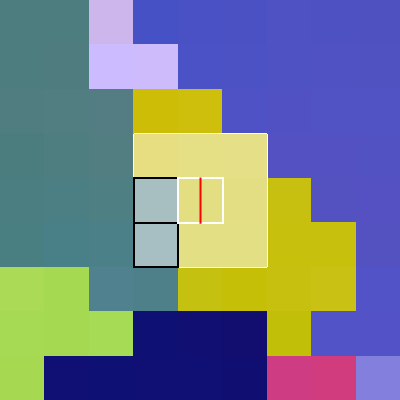}}
\hspace{0.2cm}
\subfigure[]{\includegraphics[width=4cm]{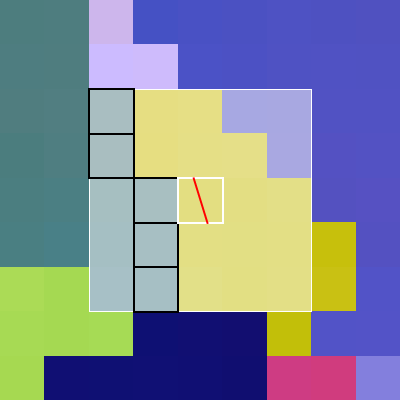}}
\hspace{0.2cm}
\subfigure[]{\includegraphics[width=4cm]{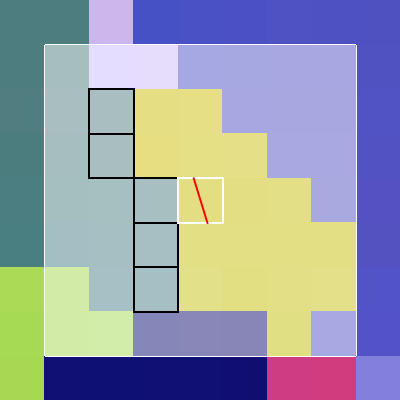}}
\caption{Enlarged section of the top plane sketching the calculation
  of the in-plane GB slope (red line) in a selected GB voxel (thick
  white square) using same-like GB voxels (thick black squares) within
  the prescribed nearest-neighbor voxel region of size $(2N_n+1)\times
  (2N_n+1)$ (thin white square) for (a) $N_n=1$, (b) $N_n=2$ and (c)
  $N_n=3$. The slope is calculated from the moment of inertia of the
  2D discrete object composed of black square centers: the slope
  corresponds to the eigenaxis with the smallest eigenvalue.}
\label{fig:methSES3}
\end{figure}

In the second step, each GB voxel is assigned an in-plane GB slope by
accounting for its neighborhood of same-like GB voxels (having the
same color pair). The centers of these neighboring same-like voxels
compose a 2D discrete object which is further used to calculate the
(in-plane) axis with the smallest moment of inertia. Such an axis is
taken to be the principal direction of the (complex) object
surrounding the corresponding GB voxel and is therefore used here to
represent also the in-plane GB slope (denoted by $k_0$). The above
step is demonstrated in Figure \ref{fig:methSES3} for three different
ranges of the neighborhood.  Here, the range is characterized by
parameter $N_n=1,2,3$ which counts the number of voxel layers
surrounding a GB voxel.

\begin{figure}[H]
  \centering
\subfigure[]{\includegraphics[width=4cm]{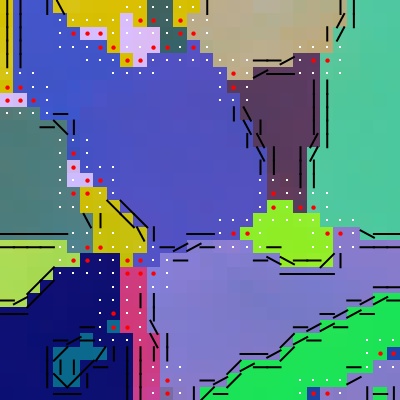}}
\hspace{0.2cm}
\subfigure[]{\includegraphics[width=4cm]{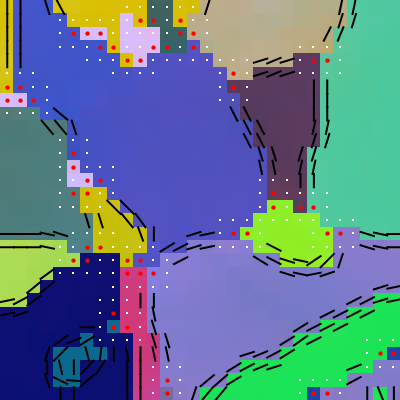}}
\hspace{0.2cm}
\subfigure[]{\includegraphics[width=4cm]{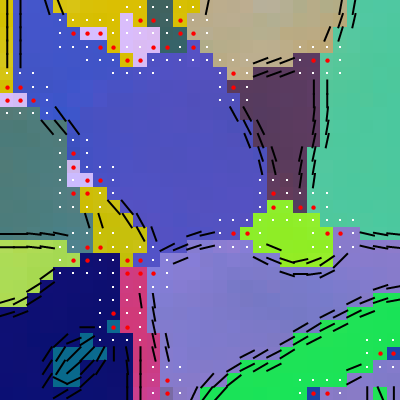}}
\caption{Detection of GB on the top plane. Black lines denote the
  calculated in-plane GB slopes of GB voxels using (a) $N_n=1$, (b)
  $N_n=2$ or (c) $N_n=3$ surrounding voxel layers. Red points denote
  triple points and white points a $3\times 3$ neighborhood of each
  triple point where GB detection is avoided.}
\label{fig:methSES4}
\end{figure}

In Fig. \ref{fig:methSES4} the in-plane GB slopes are shown for all GB
voxels using again three different ranges of the considered
neighborhoods ($N_n=1,2,3$). Depending on the grain size, the
performance of the method varies locally for a given $N_n$. Generally,
smoother slopes are produced for larger $N_n$, while better
recognition of smaller grains is found for smaller $N_n$. Here,
$N_n=2$ or 3 seems to be the optimum choice.

\begin{figure}[H]
  \centering
\subfigure[]{\includegraphics[width=4cm]{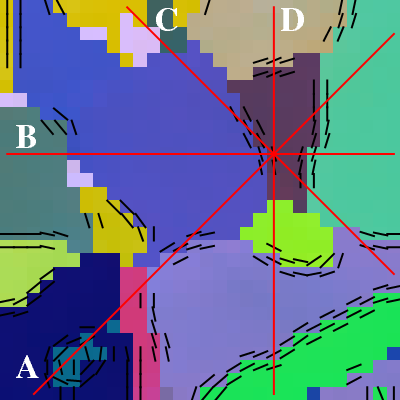}}
\hspace{1cm}
\subfigure[]{\includegraphics[width=4cm]{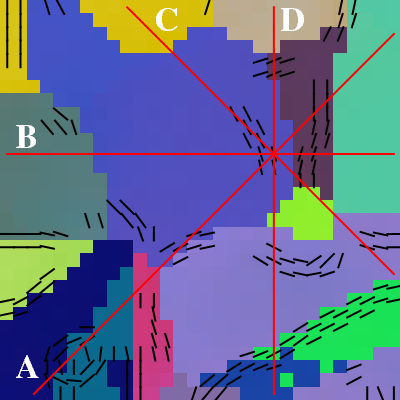}}\\
\subfigure[]{\includegraphics[height=1.cm]{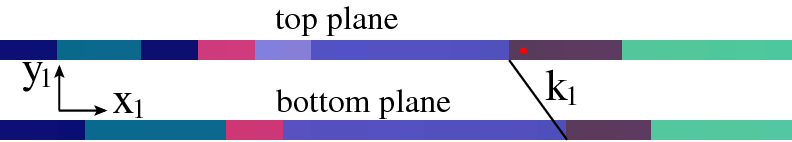}}
\hspace{1cm}
\subfigure[]{\includegraphics[height=1.cm]{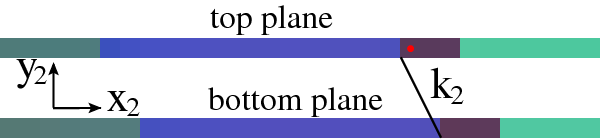}}\\
\subfigure[]{\includegraphics[height=1.cm]{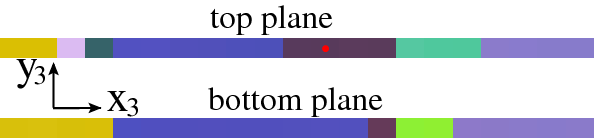}}
\hspace{1cm}
\subfigure[]{\includegraphics[height=1.cm]{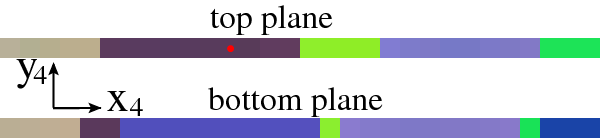}}
\caption{ (a) Top plane and (b) bottom plane with black lines denoting
  in-plane GB slopes of the top plane (for $N_n=2$) and red lines
  denoting four vertical planes A, B, C, D used to make four vertical
  cross sections shown in (c)-(f). Same GB type (color pair) is
  detected on the bottom plane in cross sections (c) A and (d) B,
  which allows to define the out-of-plane GB slopes $k_1$ and $k_2$,
  respectively.}
\label{fig:methSES5}
\end{figure}

In the third step of the method, four vertical cross sections are
produced for each GB voxel on the top plane, see Fig.
\ref{fig:methSES5}. By identifying same GB types (color pairs) on the
bottom plane, the out-of-plane GB slopes $k_i$ ($i=1,2,3,4$) can be
calculated in each cross section. It is clear, moreover, that the
accuracy of $k_i$ improves with increasing distance between the two
planes.

\begin{figure}[H]
  \centering
\subfigure[]{\includegraphics[width=3cm]{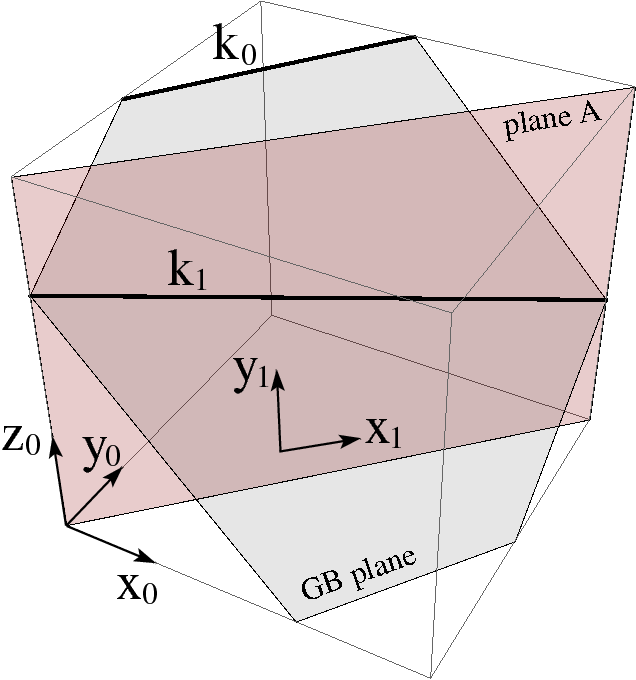}}
\hspace{0.2cm}
\subfigure[]{\includegraphics[width=3cm]{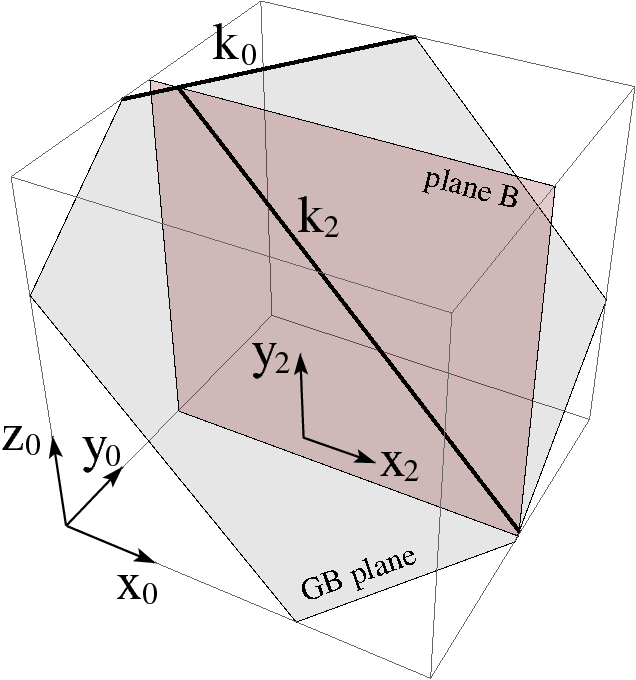}}
\hspace{0.2cm}
\subfigure[]{\includegraphics[width=3cm]{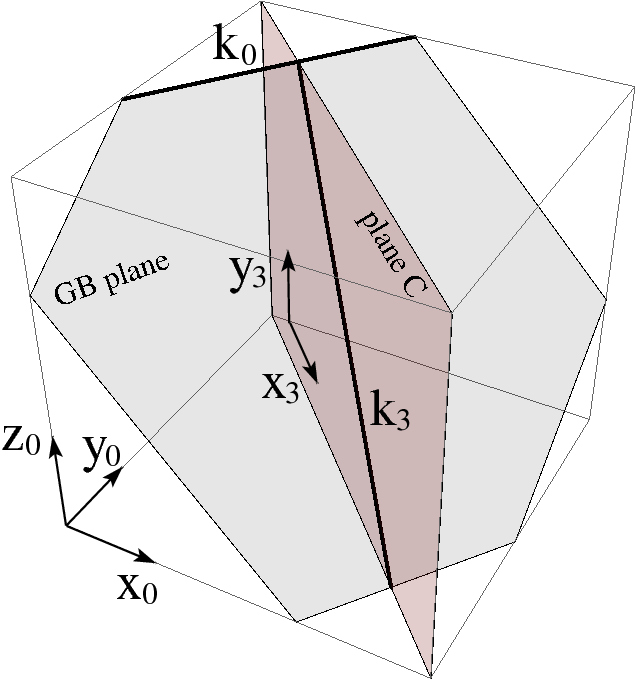}}
\hspace{0.2cm}
\subfigure[]{\includegraphics[width=3cm]{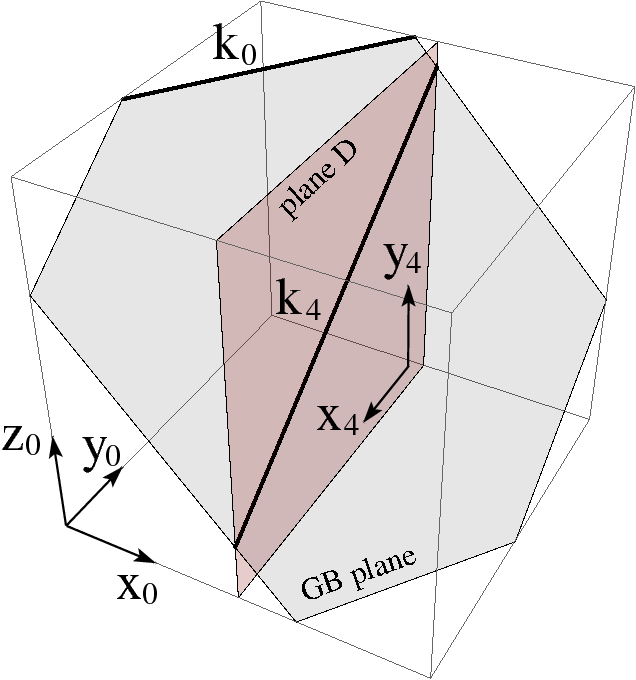}}
\caption{ Four 3D sketches showing different vertical cross sections
  (red planes) crossing the assumed GB plane (gray plane). These plots
  are used to derive the expressions for GB normal $\nn$ as a function
  of the in-plane slope $k_0$ at the top plane and one of the
  out-of-plane slopes $k_i$ ($i=1,2,3,4$), see Eq. (\ref{eq:n}).}
\label{fig:methSES6}
\end{figure}

In the last step, GB normals are finally calculated given the in-plane
and out-of-plane slopes of all the GB voxels. A GB normal $\nn$,
assigned to one particular GB voxel, is calculated from the in-plane
slope $k_0$ and out-of-plane slope $k_i$, following the definitions in
Fig. \ref{fig:methSES6}, as (omitting the normalization)
\be
  \nn\sim (-k_0, 1, \frac{k_0-1}{\sqrt{2}k_1})
      \sim (-k_0, 1, \frac{k_0}{k_2})
      \sim (-k_0, 1, \frac{k_0+1}{\sqrt{2}k_3})
      \sim (-k_0, 1, \frac{1}{k_4}).
\label{eq:n}
\ee
In principle, one $k_i$ and $k_0$ are enough to determine a GB normal
$\nn$ unambiguously. However, if more than one $k_i$ is available, an
average out-of-plane slope $\langle k_4\rangle$ is calculated first to reduce the error employed in the estimation of $k_i$. In this
respect, all available $k_i$ are first transformed to one common cross
section (labeled D) to obtain
\be
  k_{4,1}=\frac{\sqrt{2}k_1}{k_0-1},\quad
  k_{4,2}=\frac{k_2}{k_0},\quad
  k_{4,3}=\frac{\sqrt{2}k_3}{k_0+1},\quad
  k_{4,4}=k_4,
\label{eq:k4}
\ee
using Eq. (\ref{eq:n}). The average slope $\langle k_4\rangle$ is then
calculated as
\be
  \langle k_4\rangle = 
     \frac{\sum_j \sin{\left(\arctan{(k_{4,j})}\right)}}
     {\sum_j \cos{\left(\arctan{(k_{4,j})}\right)}}
     ,
\label{eq:avek4}
\ee
using the rule for calculating the mean of angles. Finally, a GB normal is
calculated as
\be
  \nn\sim (-k_0, 1, \frac{1}{\langle k_4\rangle})
\label{eq:n4}
\ee
followed by a proper normalization.\\

\begin{figure}[H]
\centering
\subfigure[]{\includegraphics[height = 3.cm]{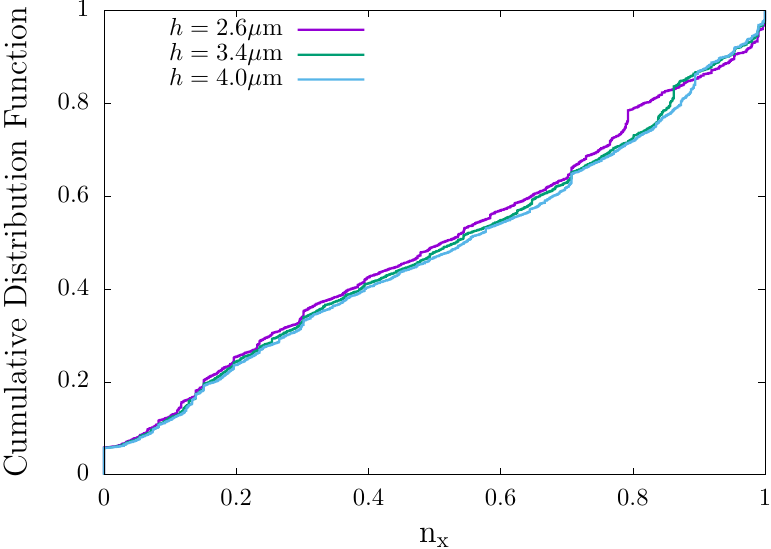}}
\hspace{0.cm}
\subfigure[]{\includegraphics[height = 3.cm]{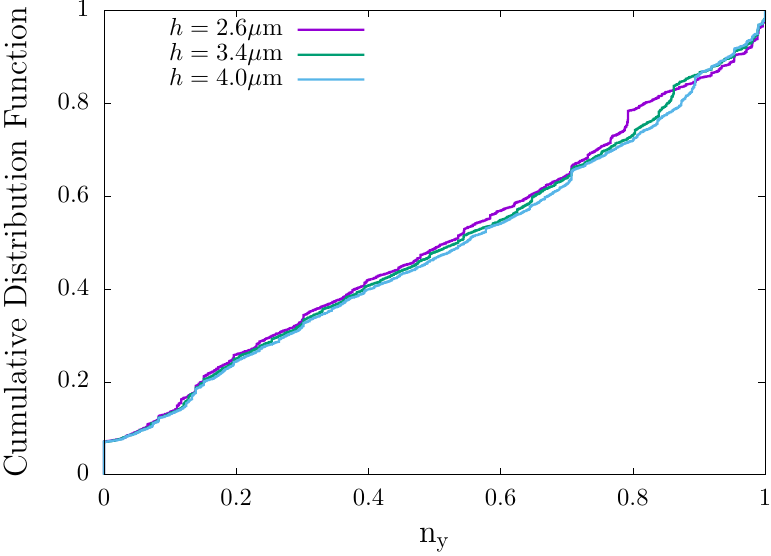}}
\hspace{0.cm}
\subfigure[]{\includegraphics[height = 3.cm]{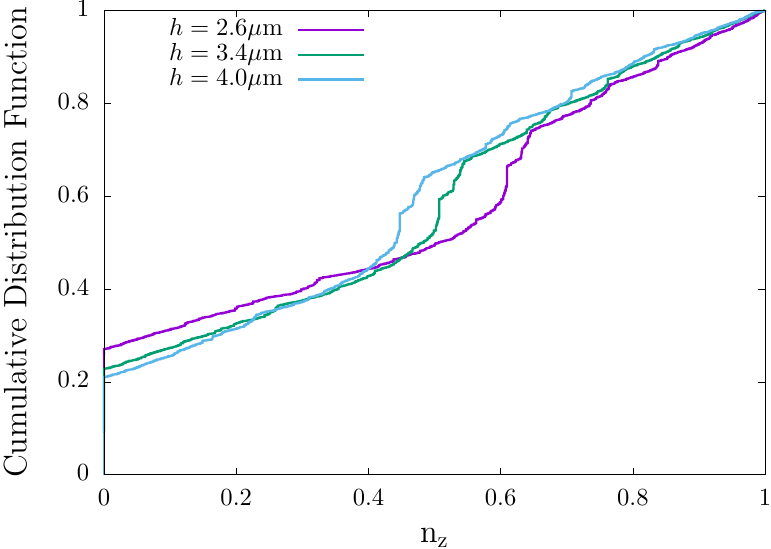}}
\caption{\textcolor{black}{Cumulative distribution functions of GB normal components as a function of the distance between the top and bottom planes $h$}}
\label{fignew1}
\end{figure}

\begin{figure}[H]
\centering
\subfigure[]{\includegraphics[height = 3.cm]{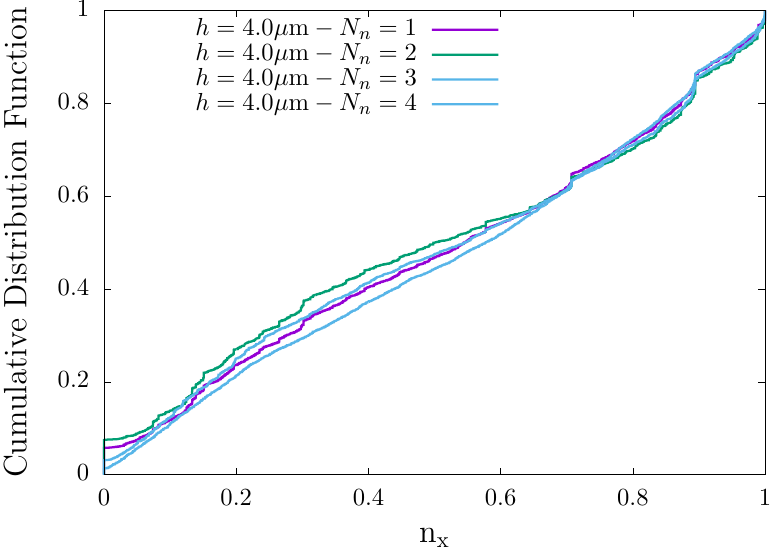}}
\hspace{0.cm}
\subfigure[]{\includegraphics[height = 3.cm]{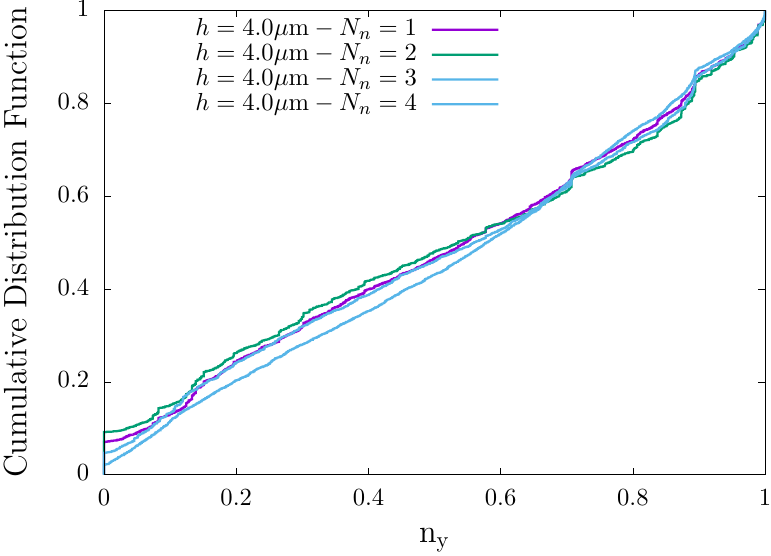}}
\hspace{0.cm}
\subfigure[]{\includegraphics[height = 3.cm]{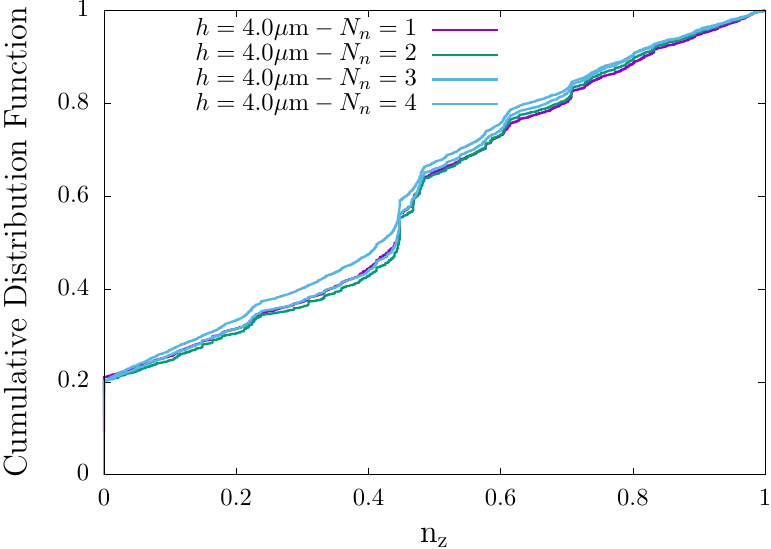}}
\caption{\textcolor{black}{Cumulative distribution functions of GB normal components as a function of the value of the smoothing parameter $N_n$}}
\label{fignew2}
\end{figure}

\textcolor{black}{The effects of varying the distance between the top and bottom planes $h$ and the smoothing parameter $N_n$ on the GB normal components are shown on Figs.~\ref{fignew1},~\ref{fignew2}. The parameter $h$ affects mostly the $n_z$ component (Fig.~\ref{fignew1}c), while the effect on the other components is weaker (Fig.~\ref{fignew1}a,b).  Due to the discrete nature of the EBSD data, the accuracy of the determination of GB out-of-plane component increases with higher values of $h$, while lower values are needed to assess GB normals close to the surface. In the following, $h = 4\mu$m is chosen as a compromise, keeping in mind that the GB normal computed corresponds to an average value over a thickness $h$.}

\textcolor{black}{The effect of the smoothing parameter $N_n$ is more limited, as shown on Fig.~\ref{fignew2}. As a compromise between the detection of small grains and smoothing of the 2D GB, $N_n = 2$ is chosen for the distributions reported in the paper.}

\begin{figure}[H]
\centering
\subfigure[]{\includegraphics[height = 4.5cm]{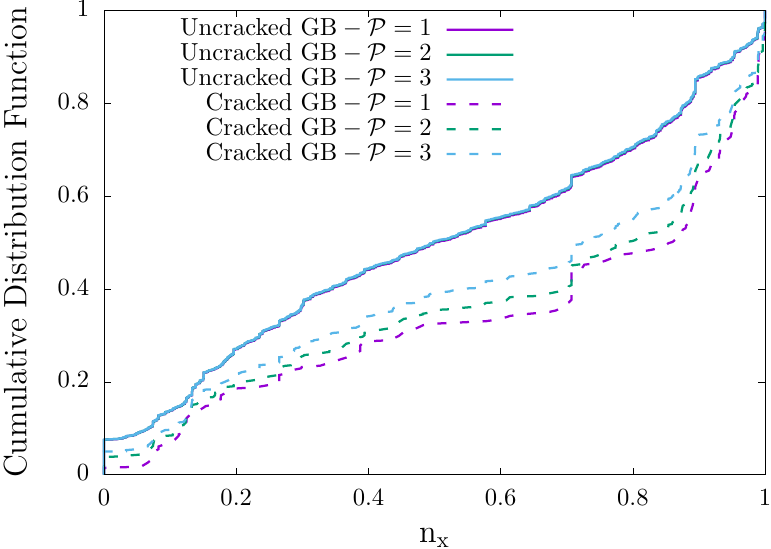}}
\hspace{0.25cm}
\subfigure[]{\includegraphics[height = 4.5cm]{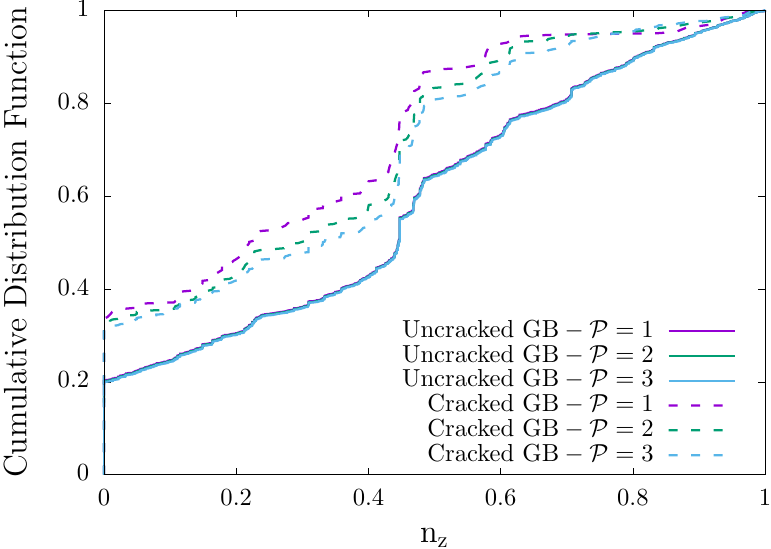}}
\subfigure[]{\includegraphics[height = 4.5cm]{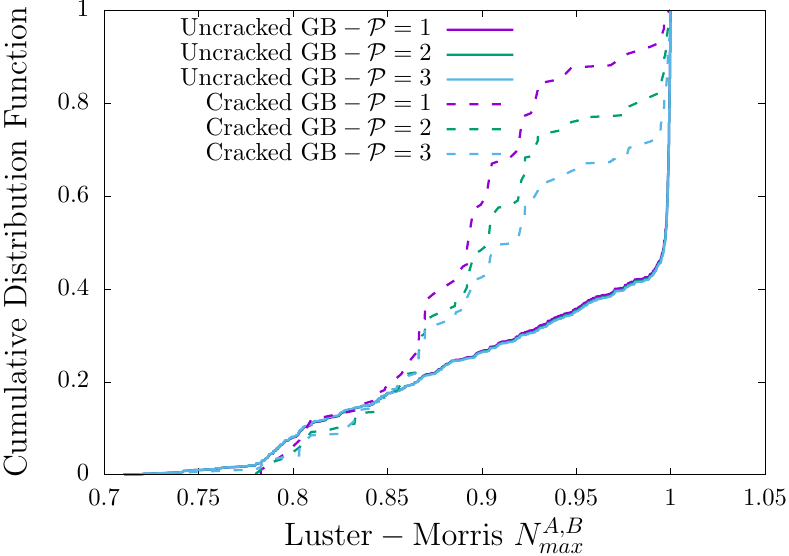}}
\hspace{0.25cm}
\subfigure[]{\includegraphics[height = 4.5cm]{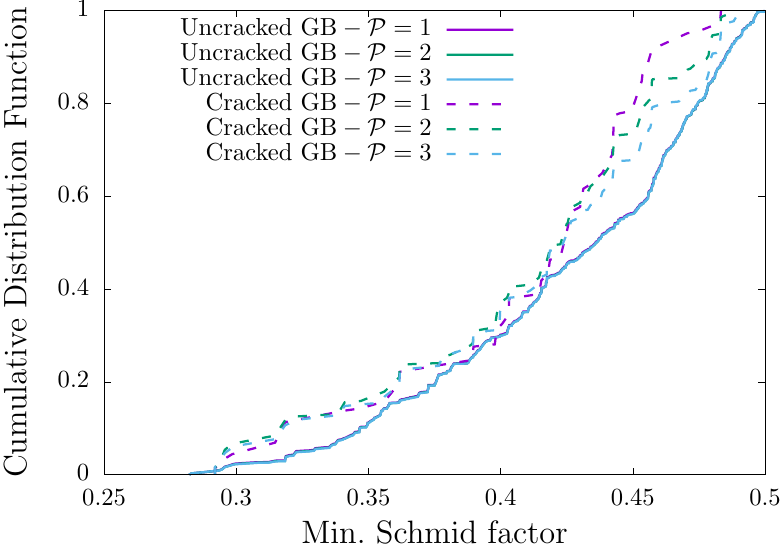}}
\caption{\textcolor{black}{Cumulative distribution functions of (a) $n_x$ (b) $n_z$ (c) Luster-Morris parameter and (d) minimal Schmid factor for uncracked and cracked GB as a function of the parameter $\mathcal{P}$ used to project cracks on GB}}
\label{fignew3}
\end{figure}

\textcolor{black}{Finally, the effect of the projection of cracks on GB is assessed in Fig.~\ref{fignew3}. As discussed in Section~\ref{incra}, a voxel at a position $\{x,y\}$ of the EBSD map is considered to correspond to a crack if $\sqrt{(x-x_c)^2 + (y - y_c)^2} \leq \mathcal{P} \Delta$, where $\{x_c,y_c\}$ is the location of a crack determined from SEM observations, $\Delta$ the step used for ESBD analysis and $\mathcal{P}$ is a parameter controlling the projection process. The effect of $\mathcal{P}$ on the cumulative distribution functions is shown on Fig.~\ref{fignew3}. Due to the relatively small number of cracks, the distributions for uncracked GB are not affected by the choice of the parameter $\mathcal{P}$. For cracked GB, higher values of $\mathcal{P}$ lead to distributions closer to the ones for uncracked GB, which is consistent with the fact that higher number of false cracked GB will be considered. The fraction of cracked GB depends on the choice of the parameter $\mathcal{P}$ and is equal to $1.6\%, 3.2\%, 4.9\%$ for $\mathcal{P} = 1, 2, 3$, respectively. In order to minimize the number of false cracked GB, the smallest value of the parameter $\mathcal{P}=1$ is considered for all \textit{cdf} presented in the main part of the manuscript. This leads to rather small fraction of cracked GB and thus to uncertainties that are evaluated by computed \textit{cdf} 95\% confidence bounds.}

\section*{Appendix C: Convergence analysis of numerical simulations}

\begin{figure}[H]
\centering
\subfigure[]{\includegraphics[height = 4.5cm]{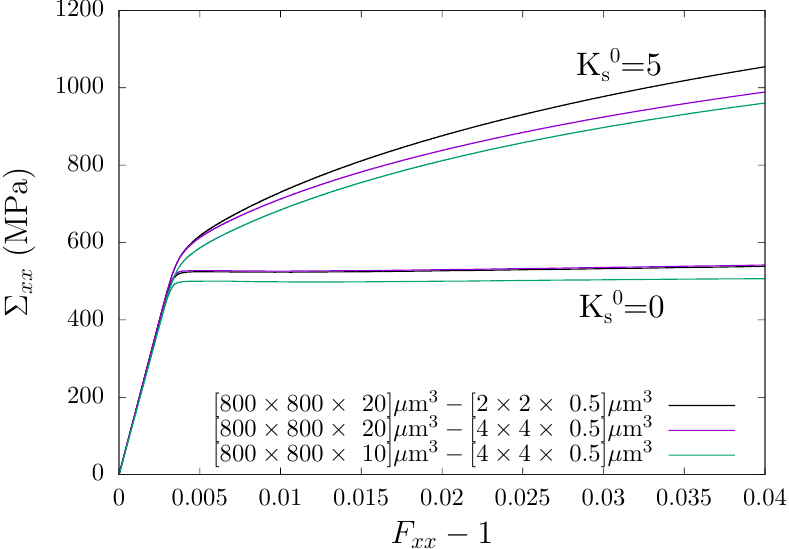}}
\hspace{0.25cm}
\subfigure[]{\includegraphics[height = 4.5cm]{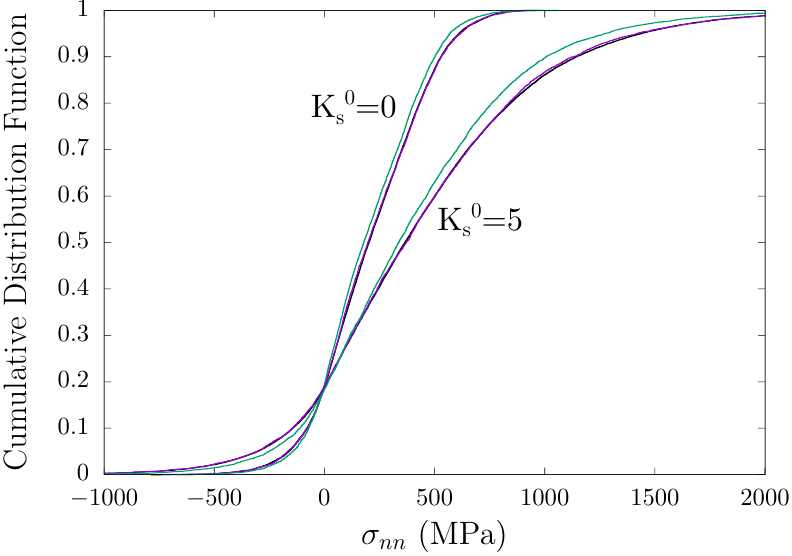}}
\caption{\textcolor{black}{(a) Aggregate stress-strain curves and (b) cumulative distribution functions of intergranular normal stresses: Effect of the aggregate size ($[800 \times 800 \times 20]\mathrm{\mu m}^3$ \textit{vs.} $[800 \times 800 \times 10]\mathrm{\mu m}^3$) and in-plane voxel size ($2\mathrm{\mu m} $ \textit{vs.} $4\mathrm{\mu m} $). The out-of-plane voxel size is set to $0.5\mathrm{\mu m} $ }}
\label{fignew4}
\end{figure}

\textcolor{black}{The convergence of the numerical results presented in Section~\ref{sec3} are assessed with respect to the aggregate size and voxel size. Figs.~\ref{fignew4}a,b show the macroscopic stress-strain curves and distributions of intergranular normal stresses computed at the free surface, for the various parameters considered. Convergence of the numerical results is already achieved for an in-plane voxel size of $4\mu$m for $K_s^0=0$, larger than the value of $2\mu$m used for all simulation results reported in Section~\ref{sec3}. A slight effect is however noticed for $K_s^0=5$. A more significant effect is observed on the stress-strain curves (Fig.~\ref{fignew4}a) between aggregates of thickness $20\mu$m and $10\mu$m, coming from the influence of the free surface. The key point is that the influence of the aggregate's thickness (and in-plane voxel size) on the intergranular normal stresses at the surface is weak, as shown on Fig.~\ref{fignew4}b, justifying the convergence of the results reported in Section~\ref{sec3}.}

\bibliography{biblio}

\end{document}